\documentclass[12pt]{article}

\usepackage{amsmath,amsthm} 
\usepackage{amssymb,mathrsfs} 
\usepackage{a4wide} 
\usepackage{graphicx}
\usepackage{physics}
\usepackage{color,subfigure} 
\usepackage{enumerate}
\usepackage[normalem]{ulem}
\usepackage{cancel}
\usepackage{bbm}
\usepackage{tikz}
\usepackage{hyperref}
\usepackage{nccmath}
\usepackage{mathtools}

\usetikzlibrary{arrows}

\mathtoolsset{showonlyrefs=true}

\newtheorem{remark}{Remark}

\DeclareMathAlphabet{\mathpzc}{OT1}{pzc}{m}{it}

\renewcommand{\d}{\mathrm{d}}
\newcommand{\cL}{\mathcal{L}}
\newcommand{\1}{\mathbbm{1}}
\newcommand{\Dt}{{\Delta t}}
\newcommand{\iid}{i.i.d.}

\DeclareMathOperator*{\argmin}{argmin}

\newcommand{\E}{\mathbb{E}}

\newcommand{\Yr}{Y^r}
\newcommand{\Ybr}{Y^{\mathbf{r}}}
\newcommand{\Rr}{\mathcal{R}^r}

\newcommand{\Lambdar}{\Lambda^r}

\newcommand{\e}{\mathrm{e}}
\newcommand{\R}{\mathbb{R}}
\newcommand{\Var}{\mathrm{Var}}
\newcommand{\Cov}{\mathrm{Cov}}
\renewcommand{\leq}{\leqslant}
\renewcommand{\geq}{\geqslant}

\newcommand{\ct}{\alpha}

\begin{document}

\title{Fixing the flux: A dual approach to computing transport coefficients}
\author{N. Blassel$^{1,2}$ and G. Stoltz$^{1,2}$ \\
   \small 1: CERMICS, \'Ecole des Ponts, Marne-la-Vallée, France  \\
   \small 2: MATHERIALS project-team, Inria Paris, France
}

\maketitle

\abstract{We present a method to compute transport coefficients in molecular dynamics. Transport coefficients quantify the linear dependencies of fluxes in non-equilibrium systems subject to small external forcings. Whereas standard non-equilibrium approaches fix the forcing and measure the average flux induced in the system driven out of equilibrium, a dual philosophy consists in fixing the value of the flux, and measuring the average magnitude of the forcing needed to induce it. A deterministic version of this approach, named Norton dynamics, was studied in the 1980s by Evans and Morris. In this work, we introduce a stochastic version of this method, first developing a general formal theory for a broad class of diffusion processes, and then specializing it to underdamped Langevin dynamics, which are commonly used for molecular dynamics simulations.
We provide numerical evidence that the stochastic Norton method provides an equivalent measure of the linear response, and in fact demonstrate that this equivalence extends well beyond the linear response regime.
This work raises many intriguing questions, both from the theoretical and the numerical perspectives.}

\section{Introduction}\label{intro}
Molecular dynamics (MD), much like computational statistical physics at large, aims at predicting macroscopic properties of a molecular medium by the use of computer simulations. In order to do so, a classical model of the interactions between atoms is constructed, and dynamical evolution laws are specified.
One can then simulate typical trajectories of the system through phase space, which,  provided they are long enough, allows to estimate to a prescribed level of statistical accuracy the quantities of interest. See~\cite{b91} for a general theoretical account of statistical physics,~\cite{t10,at17,fs01} for an emphasis on numerical methods, and~\cite{lm15} for an overview of the mathematical aspects of molecular simulation.
MD simulations may prove useful when the properties of interest are impractical to measure due to physical or cost constraints associated with the experimental setup, or alternatively serve as surrogate tests for theoretical models. At any rate, MD has grown to occupy an important role in many applications ranging from pharmacology and molecular biology to materials science and condensed matter physics, besides having had a significant impact on statistical methodology at large, through the dissemination of tools such as the Metropolis--Hastings algorithm~\cite{m53}.
First applications of MD included the computation of static properties at equilibrium, in particular thermodynamic quantities or free energy differences, which still are of considerable interest today. We refer to~\cite{cks05} for a historical perspective.

Another, more difficult problem is the measurement of dynamical properties, quantities which depend on the trajectory itself, and capture the behavior of systems undergoing a macroscopic time evolution: this is the object of non-equilibrium statistical physics.
Of particular interest is the computation of transport coefficients, giving a measure of the sensitivity in the response of a system at equilibrium to the magnitude of perturbations driving it out of equilibrium. See~\cite{em08,TD17} for overviews, and~\cite[Section 5]{ls16} for a mathematical presentation.
Transport coefficients are of practical importance, since they characterize the diffusion, heat conduction or viscosity properties of a molecular medium, and enter as parameters in macroscopic evolution equations such as the Navier--Stokes equation.
One standard approach to compute transport coefficients by molecular dynamics relies on the celebrated Green--Kubo formula~\cite{g54}, which expresses transport coefficients as integrated time-correlations between appropriate fluxes in the system at equilibrium. Another standard approach, see for instance~\cite{cj75}, is to directly simulate the non-equilibrium perturbation, and to measure the resulting average response, which is, at the macroscopic level and in the limit of a small perturbation, proportional to the magnitude of this perturbation: this is the so-called non-equilibrium molecular dynamics (NEMD) approach, reviewed in~\cite{cks05}. Estimators deriving from these approaches however suffer from large statistical errors, as quantified in~\cite[Proposition 2.4]{ss23} for instance, and convergence requires the simulation of very long trajectories, which comes at a high computational cost. A key metric to measure this cost is, for any given method, the asymptotic variance of estimators of the transport coefficient. Although some variance reduction techniques have been proposed to compute transport coefficients (see~\cite{s22} for a recent review), efficiently estimating these quantities is are still an important area of research.

From a thermodynamic point of view, the NEMD approach can be understood as fixing the magnitude of the non-equilibrium forcing, and measuring the resulting flux in the system driven out of equilibrium, which is conceived as a microscopic state variable. For a small enough magnitude of the forcing, this flux responds approximately linearly to the forcing, and the transport coefficient is precisely the associated proportionality constant. This method also yields a construction of the non-equilibrium ensemble, by defining it as the steady state of the system evolving according to the microscopic dynamics.

However convenient from an implementation standpoint, there is no physical reason to a priori favor an ensemble in which the forcing field is exactly fixed. At the macroscopic level, fluxes and forces play a symmetric role, which opens the possibility for dual approaches in the computation of non-equilibrium responses. Other approaches have been proposed to construct non-equilibrium ensembles in which the forcing does not play such a distinguished role. One can cite McLennan ensembles~\cite{mn10}, expressing non-equilibrium steady-states as first-order perturbations of the Boltzmann distribution, where the leading order correction term is the time-integral of some conjugate response; or the dynamical approaches by Komorowski and Olla~\cite{klos22}, whereby an average flux is induced by a time-periodic forcing. When considering a dual perspective, the most radical approach is to fix the flux exactly, and measure the average magnitude of the forcing needed to induce it in the microscopic dynamics. This yields an alternative measure of the transport coefficient. By analogy with Ohm's law, one can think of the NEMD approach as a way to measure a conductance, and the radical dual approach as a way to measure a resistance.

The microscopic implementation of this dual approach is what we will refer to as the Norton method, in accordance with the terminology proposed by Evans and Morriss in~\cite{em85}, by reference to the Norton--Th\'evenin reciprocity from electrical circuit theory. This approach amounts to considering a constrained dynamics on a submanifold of phase space consisting of a level-set of the flux observable. The idea of using constrained dynamics to simulate non-equilibrium systems already appeared in the 1983 work~\cite{ehfml83}, where it was applied to capture the mobility,
and was also explored in~\cite{hpc93}. It was further applied to shear viscosity computations in~\cite{ee86}, where the consistency of the approach was demonstrated. From a theoretical perspective, formal results under ergodic hypotheses were obtained, including linear response expressions for the transport coefficient in~\cite{em85}, as well as a result on the equivalence of non-equilibrium ensembles in~\cite{e93}.

However, Norton dynamics were only considered in a deterministic setting, and, despite promising results, their potential for practical use was not fully explored. In particular, the numerical performance of estimators of transport coefficients based on time averages of this dynamics has not, to the best of our knowledge, been studied yet. Our aim is to extend the Norton philosophy to the stochastic setting, both for academic motivations (obtaining new results on equivalence of ensembles in non-equilibrium systems), and for numerical reasons, as Norton dynamics potentially allow to more efficiently compute transport coefficients. We will consider both general diffusion processes, and Langevin-type dynamics, which are commonly used in molecular dynamics~\cite{PavliotisBook}.

\paragraph{Contributions of this work.} Let us highlight our main contributions:
\begin{itemize}
\item{We define a stochastic version of the Norton method, which is convenient from the theoretical point of view as rigorous ergodicity results exist for stochastic dynamics, and are also relevant for simulation since stochastic dynamics are nowadays very commonly used in molecular dynamics.}
\item{We specialize the Norton method to underdamped Langevin dynamics, and apply it to compute the mobility and shear viscosity of a fluid. We demonstrate numerically that the Norton method gives consistent estimations of the linear response in the thermodynamic limit.}
\item{We observe on the numerical examples we consider that the non-equilibrium responses coincide in fact far outside of the linear regime, raising the question of equivalence between non-equilibrium ensembles.}
\item{We offer numerical evidence that, in some situations, the Norton method is preferable to the standard NEMD approach, in the sense that the Norton estimators for the transport coefficients lead to estimates with a smaller statistical error than their NEMD counterparts.}
\end{itemize}
Many points in the mathematical analysis of the Norton approach are left open at this stage. This should be seen as an invitation to further study the properties of these intriguing dynamics

\paragraph{Outline.}
This paper is organized as follows. We recall in Section~\ref{sec:nemd} a general framework for stochastic dynamics out of equilibrium. In Section~\ref{sec:norton}, we introduce the Norton approach, deriving an expression for the diffusion process defining the dynamics, before generalizing the approach to the case of multiple constraints and time-dependent fluxes. In Section~\ref{sec:non_eq}, we specialize the setting to non-equilibrium kinetic Langevin dynamics used for mobility and shear viscosity computations, describing how to apply the Norton philosophy. We also give a physical interpretation of the Norton dynamics, as one satisfying an oblique version of Gauss's principle of least constraint. In Section~\ref{sec:discr}, we discuss how to discretize stochastic Norton dynamics, describing in particular a family of schemes obtained by an operator splitting approach. We present the results of our numerical experiments in Section~\ref{sec:numerical}, demonstrating the consistency with usual NEMD in the thermodynamic limit, and motivating that Norton simulations can be more efficient than NEMD ones. We conclude in Section~\ref{sec:perspective} by discussing the many open questions and perspectives raised by the Norton method in the stochastic context.

\section{Non-equilibrium molecular dynamics}\label{sec:nemd}
We recall in this section the standard framework of NEMD, first presenting the reference dynamics, before introducing the non-equilibrium dynamics and defining transport coefficients. We finally discuss the statistical properties of NEMD estimators of the transport coefficient. We refer the interested reader to~\cite[Section 5]{ls16} and~\cite{ss23} for a more in-depth discussion of the mathematical properties of the NEMD method.
\paragraph{Reference dynamics.}\label{par:ref}
We consider a class of non-equilibrium ensembles, obtained as the steady-states of certain stochastic processes corresponding to a perturbation of a reference process (usually a dynamics at equilibrium).
 We denote by~$\mathcal X$ the state space of the system.
Typically, we consider~$\mathcal X= \mathbb{R}^d$ or~$\mathbb{T}^d$, with~$\mathbb{T}= \mathbb{R}/\mathbb{Z}$ the one-dimensional torus, or a product of the two. We introduce a smooth vector field~$b: \mathcal{X}\to \mathbb{R}^d$, corresponding to the equilibrium drift, and a matrix valued map~$\sigma : \mathcal X\to \R^{d\times m}$ corresponding to the diffusion.
The reference dynamics is the following stochastic differential equation (SDE):
\begin{equation}
    \d X_t = b(X_t)\,\d t +\sigma(X_t)\,\d W_t,
\end{equation}
where~$(W_t)_{t\geq 0}$ a standard~$m$-dimensional Brownian motion.
Common choices include the overdamped Langevin dynamics
\[\d X_t = -\nabla V(X_t)\,\d t + \sqrt{\frac{2}{\beta}}\,\d W_t,\]
and the underdamped or kinetic Langevin dynamics (see Equation~\eqref{eq:langevin_equation} below).

\paragraph{Non-equilibrium perturbations.}\label{par:neq}
We consider in this work the case where the perturbation arises from an external non-gradient forcing in the drift of the underlying diffusion process, determined by a smooth vector field~$F:\mathcal{X}\to \R^d$.
The non-equilibrium dynamics is given by the following SDE:
\begin{equation}
    \label{eq:thevenin_dynamics}
    \d X^\eta_t = (b+\eta F)(X^\eta_t)\,\d t +\sigma(X^\eta_t)\,\d W_t.
\end{equation}
The parameter~$\eta\in\mathbb R$ is a scalar modulating the strength of the perturbation.
The equilibrium dynamics is recovered in the absence of a non-equilibrium forcing, i.e.~$\eta=0$.
The infinitesimal generator of the dynamics~\eqref{eq:thevenin_dynamics} can be decomposed as the sum
\begin{equation}
    \label{eq:thevenin_generator_nemd}\cL_\eta=\cL+\eta \widetilde{\cL},\qquad \cL=b \cdot \nabla + \frac12\sigma  \sigma^\intercal : \nabla ^2,\qquad \widetilde{\cL}=F\cdot \nabla,
\end{equation}
where~$\nabla^2$ denotes the Hessian matrix and~$:$ denotes the Frobenius inner product~$A:B=\mathrm{Tr}(A^\intercal B)$. 
Note that~$\cL$ is the generator of the reference dynamics, and~$\widetilde{\cL}$ encodes its perturbation.
The invariant probability measure satisfies the stationary Fokker-Planck equation
\begin{equation}
    \label{eq:thevenin_fokker-planck equation}
    \cL_\eta^{\dagger} \psi_\eta = 0,
\end{equation}
where~$\cL_\eta^{\dagger}$ is the flat~$L^2(\mathcal X)$-adjoint of the generator. Existence and regularity results for solutions of~\eqref{eq:thevenin_fokker-planck equation} depend on the particular form of the dynamics, as do properties pertaining to convergence to the steady-state. It is often possible to leverage the standard analytical framework of parabolic and elliptic partial differential equations to prove this convergence, although for degenerate diffusions such as the underdamped Langevin dynamics, one has to resort to more sophisticated tools, such as hypoellipticity~\cite{h67} to obtain regularity of the steady-state, or hypocoercivity~\cite{v06} to show exponential decay of the evolution semigroup.

We assume in the remainder of this section that, for~$|\eta|$ small enough, the dynamics admits a unique invariant probability measure for which it is ergodic, and denote the corresponding expectation by~$\mathbb{E}_\eta$. Given a response observable~$R : \mathcal X \to \mathbb R$ such that
\begin{equation}
\label{eq:R_centered_Thevenin}
   \mathbb{E}_0\left[ R\right]=0, 
\end{equation} which we think of as measuring a flux in the system out of equilibrium, we define the associated transport coefficient as the following limit, provided it is well defined:
\begin{equation}
    \label{eq:def_thevenin_transport_coeff}
    \ct = \underset{\eta\to 0}{\lim}\frac{\mathbb{E}_{\eta}\left[R\right]}{\eta}.
\end{equation}
Rigorous assumptions under which this limit exists are given in~\cite{h10}. This definition suggests a simple and natural method to estimate these coefficients: one can compute ergodic averages of~$R$ over trajectories of the non-equilibrium dynamics~\eqref{eq:thevenin_dynamics}, and estimate the linear relation between~$\eta$ and~$R$ for one or several values of~$\eta$ in the linear response regime. The finite-difference estimator for the limit~\eqref{eq:def_thevenin_transport_coeff} is given by the following ergodic average:
\begin{equation}
    \label{eq:thevenin_rho_estimators_continuous_time}
    \widehat{\ct}_{T,\eta}=\frac1{\eta T}\int_0^T R(X^\eta_t)\,\d t.
\end{equation}
The consistency of such estimators is a consequence of the pathwise ergodicity for the process~\eqref{eq:thevenin_dynamics}, which in many cases can be proven using the results of~\cite{k87}. The latter result also implies the uniqueness of the steady-state. The existence of the steady-state is often obtained by Lyapunov techniques, see for instance~\cite[Theorem~8.3]{rb06}.

\paragraph{Statistical properties of the estimator~$\widehat{\ct}_{T,\eta}$.}
A challenge posed by the NEMD method is that the estimator~\eqref{eq:thevenin_rho_estimators_continuous_time} is plagued by large statistical errors when~$|\eta|$ is small, which is often required to remain in the linear response regime, and avoid biases arising from nonlinear terms in the response. More precisely, one can show that the asymptotic variance of the estimator~\eqref{eq:thevenin_rho_estimators_continuous_time} scales as~$\eta^{-2}$ as~$\eta$ approaches~$0$. Indeed, under technical decay conditions on the evolution semigroup generated by~\eqref{eq:thevenin_generator_nemd}, one can easily show that the asymptotic variance is given by
\begin{equation}
\label{eq:av_thevenin}
\sigma_{\eta}^2 = \lim_{T\to\infty}\, T\Var_\eta\left(\widehat{\ct}_{T,\eta}\right) =  \frac{2}{\eta^2}\int_0^\infty \E_{\eta}[R(X_{t}^\eta)R(X_0^\eta)]\,\d t,
\end{equation}
where~$\Var_\eta$ denotes the variance with respect to~$\E_\eta$.
Defining the correlation time by
\begin{equation}
\label{eq:nemd_correlation_time}
    \Theta_{\eta}(R) = \int_{0}^\infty \frac{\E_\eta[R(X_t^\eta)R(X_0^\eta)]}{\E_\eta[R^2]}\,\d t,
\end{equation}
we further get, using a first-order expansion in powers of~$\eta$ (whose validity has to be verified on a case-by-case basis), 
\begin{equation}
    \label{eq:nemd_av_asymptotic}
    \sigma_{\ct,\eta}^2=\frac{2}{\eta^2}\Var_\eta(R)\Theta_{\eta}(R)=\frac2{\eta^2}\Var_0(R)\Theta_0(R) +\mathrm{O}\left(\frac1\eta\right).
\end{equation}
In summary, the asymptotic variance is, at dominant order in~$|\eta|$, the asymptotic variance of the time averages of~$R$ under the equlibrium dynamics, divided by~$\eta^2$.

The leading order of the asymptotic variance highlights why it is often computationally expensive to obtain accurate estimates of transport coefficients. Although the computation of such coefficients is recognized as being a difficult problem in practice, only a handful of research works have been proposed to alleviate this issue, including replacing the forcing~$F$ by a so-called synthetic forcing devised to induce the same transport coefficient while the range of the linear response ~\cite{ss23}, or relying on carefully constructed couplings~\cite{DS23}. We refer the reader to~\cite[Section~4]{s22} for a recent overview of current variance reduction techniques in non-equilibrium molecular dynamics.

\section{A stochastic Norton method}\label{sec:norton}
The Norton approach exploits the macroscopic duality between thermodynamic forces and fluxes: 
at the macroscopic level, one can equivalently choose to measure the current induced by a constant force, or the resistance opposed to a constant current.
The microscopic translation of this duality is the introduction of a new non-equilibrium ensemble in which the flux is held fixed. As in the NEMD case, we define this ensemble as the invariant probability measure for a particular stochastic process, which we refer to as the Norton dynamics.

In Section~\ref{subsec:norton_presentation}, we present the Norton pertubation approach for a generic reference dynamics of the form~\eqref{eq:thevenin_dynamics}. We then proceed in Section~\ref{subsec:norton_lambda} to write the dynamics in closed form, by making explicit the constraining force on the flux. In Section~\ref{subsec:norton_tc}, we give the expression of the Norton analogs of the transport coefficient~\eqref{eq:def_thevenin_transport_coeff}, and discuss how their statistical properties can be formally analyzed. We finally show in Section~\ref{subsec:norton_generalization} how the Norton approach can be extended to the cases of multiple forcings, or time-dependent flux constraints.

\subsection{Presentation of the dynamics}\label{subsec:norton_presentation}
At the dynamical level, the Norton ensemble is defined as the invariant probability measure of the following stochastic differential equation:
\begin{equation}
    \label{eq:norton_dynamics}
    \left\{\begin{aligned}
    \d \Yr_t &= b(\Yr_t)\, \d t + \sigma(\Yr_t)\, \d W_t +  F(\Yr_t)\,\d\Lambdar_t,\\
    R(\Yr_t) &= R(\Yr_0)=r.
    \end{aligned}\right.
\end{equation}
Here, the evolution of the state is given by the dynamics of~$\Yr_t \in \mathcal X$, and~$r\in \mathbb{R}$ is the magnitude of the response flux, which is maintained constant. The stochastic dynamics therefore evolves on the submanifold
\begin{equation}
    \label{eq:norton_submanifold}
    \Sigma_r =\left\{y\in \mathcal{X},\ R(y)=r\right\} = R^{-1}\{r\}
\end{equation}
of the full state space. The dynamics~\eqref{eq:norton_dynamics} can formally still be considered as a perturbation of the equilibrium dynamics, in the same direction as the Th\'evenin process~\eqref{eq:thevenin_dynamics}, but with a random intensity given by the stochastic process~$\Lambdar_t$, acting as the control by which the constant-flux condition is enforced. Provided~$\E_0[R]=0$, we can further interpret the Norton dynamics~$Y_t^0 \in \Sigma_0$ as an equilibrium dynamics, constrained to exactly preserve the flux. The relationship between the equilibrium ensemble in which the response fluctuates, and the Norton equilibrium ensemble in which it is exactly fixed at zero, is reminiscent of the relationship between the canonical and microcanonical ensembles when~$R$ is a spatial average over local quantities.

We next to show that~$\Lambdar$ can in fact be defined as an Itô process adapted to the natural filtration of the~$m$-dimensional Brownian motion~$W$. More precisely, we decompose the intensity of the forcing using the following ansatz:
\begin{equation}
    \label{eq:norton_lambda_sde}
    \Lambdar_t = \Lambdar_0 + \int_0^t \lambda(Y^r_s)\, \d s + \widetilde{\Lambda}_t^r,\qquad  \widetilde{\Lambda}_t^r = \int_0^t \widetilde{\lambda}(Y^r_s)\,\d W_s,
\end{equation}
where~$\lambda,\widetilde{\lambda}$ are functions defined on~$\mathcal X$, with~$\lambda$ taking values in~$\R$ and~$\widetilde \lambda$ in~$\R^{1\times m}$. This ansatz is natural for constrained dynamics (see for instance~\cite[Chapter 3]{lrs10}), and will be confirmed a posteriori (see in particular~\eqref{eq:norton_martingale_part} and~\eqref{eq:norton_lambda_expr} in Section~\ref{subsec:norton_lambda} below). The average forcing in the Norton ensemble can then be defined as the expectation of~$\lambda$ under the steady-state probability measure, neglecting the zero-mean contribution of~$\widetilde{\lambda}$ (see Section~\ref{subsec:norton_tc}).
Numerically, these averages can be computed as ergodic averages over discretized trajectories of the Norton dynamics, as discussed in Section~\ref{subsec:discr_norton}.

\subsection{A closed form for the forcing process}\label{subsec:norton_lambda}
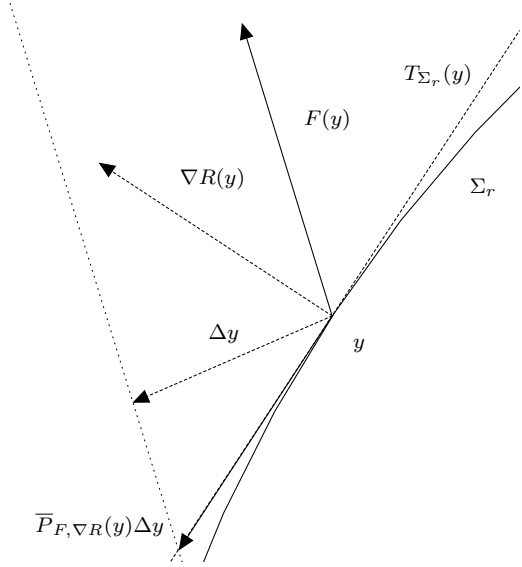
\begin{figure}
\label{fig:projector}
\centering
    \begin{tikzpicture}[line cap=round,line join=round,>=triangle 45,scale=1.5]
                    \clip(2.24,1.18) rectangle (6.82,6.13);
                    \draw [shift={(16.02,-3.77)}] plot[domain=1.54:3.42,variable=\t]({1*12.99*cos(\t r)+0*12.99*sin(\t r)},{0*12.99*cos(\t r)+1*12.99*sin(\t r)});
                    \draw [line width=0.4pt,dash pattern=on 1pt off 1pt,domain=2.24:6.82] plot(\x,{(-32.07--10.86*\x)/7.13});
                    \draw [->,dash pattern=on 1pt off 1pt] (5.16,3.36) -- (3.09,4.72);
                    \draw [->,line width=0.4pt] (5.16,3.36) -- (4.36,5.96);
                    \draw [->,line width=0.4pt,dash pattern=on 1pt off 1pt] (5.16,3.36) -- (3.39,2.59);
                    \draw [dotted,domain=2.24:6.82] plot(\x,{(-10.9--2.6*\x)/-0.8});
                    \draw [->,line width=0.4pt] (5.16,3.36) -- (3.8,1.28);
                    \begin{scriptsize}
                    \draw (5.4,3.1) node[circle] {$y$};
                    \draw (6.5,4.5) node {$\Sigma_r$};
                    \draw (6.1,5.5) node {$T_{\Sigma_r}(y)$};
                    \draw (4.1,4.58) node {$\nabla R(y)$};
                    \draw (5.12,5.1) node {$F(y)$};
                    \draw (4.2,3.2) node {$\Delta y$};
                    \draw (3.1,1.5) node {$\overline{P}_{F,\nabla R}(y)\Delta y$};
                    \end{scriptsize}
    \end{tikzpicture}
    \caption{Action of the projector~$\overline{P}_{F,\nabla R}$: the increment~$\Delta y$ is projected onto the tangent space~$T_{\Sigma_r}(y)=\{z\in \mathbb R^d\, |\, \nabla R(y)\cdot z = 0\}$ in the direction~$F(y)$.}
\end{figure}

We make precise here the expressions of the function~$\lambda(\Yr_t)$ and the martingale~$\widetilde{\Lambda}^r_t$  in~\eqref{eq:norton_lambda_sde}, which allows us to write the Norton dynamics without explicit reference to the forcing. 
We assume that~$\Lambdar$ is of the form~\eqref{eq:norton_lambda_sde}, and verify a posteriori that this ansatz is valid.
Applying Itô's formula to the constant response condition~$R(\Yr_t)=r$ yields
\begin{equation}
\label{eq:norton_ito_formula}
    \nabla R(\Yr_t)\cdot \left[b(\Yr_t)\, \d t + \sigma(\Yr_t)\, \d W_t +  F(\Yr_t)\,\d\Lambdar_t\right] +\frac 12 \nabla ^2 R(\Yr_t) : \d\left\langle M^r\right\rangle_t = 0,
\end{equation}
where~$\langle M \rangle_t$ denotes the quadratic covariation process for the martingale part in the Itô decomposition of~$\Yr$:
\begin{equation}
    \label{eq:norton_martingale_part}
    \d M^r_t = \sigma(\Yr_t)\,\d W_t + F(\Yr_t)\,\widetilde{\lambda}(Y^r_t)\,\d W_t.
\end{equation}
Using the uniqueness of the Itô decomposition, we can identify the martingale increment in~\eqref{eq:norton_ito_formula} as
\begin{equation}
    \label{eq:norton_forcing_martingale_part}
    \d \widetilde{\Lambda}^r_t = -\frac{\nabla R(\Yr_t) \cdot \sigma(\Yr_t)\,\d W_t}{\nabla R(\Yr_t)\cdot F(\Yr_t)},
\end{equation}
provided that~$\nabla R \cdot F \neq 0$ almost surely, which we assume here and in the sequel.
Plugging this equality in~\eqref{eq:norton_martingale_part} in turn gives
\[
M_t^r = \int_0^t \left(\mathrm{Id} -\frac{F(\Yr_s) \otimes \nabla R(\Yr_s)}{F(\Yr_s) \cdot \nabla R(\Yr_s)}\right)\sigma(\Yr_s)\,\d W_s = \int_0^t\overline{P}_{F,\nabla R}\sigma(Y^r_s)\,\d W_s,
\]
so that the covariation of the martingale increment is
\begin{equation}
    \label{eq:norton_covariation_increment}
    d\left\langle M^r\right\rangle_t = \left[\overline{P}_{F,\nabla R} \sigma\sigma ^\intercal \overline{P}_{\nabla F,R}^\intercal\right](\Yr_t)\,\d t.
\end{equation}
In the latter two expressions, we make use of the following non-orthogonal projector-valued maps, defined for vector fields~$A,B$ such that~$A(x)\cdot B(x)\neq 0$:
\begin{equation}
    \label{eq:norton_proj_AB}
    P_{A,B}(x) = \frac{A(x) \otimes B(x)}{A(x) \cdot B(x)},\qquad \overline{P}_{A,B}(x)=\mathrm{Id}-P_{A,B}(x).
\end{equation}
The action of the projector is given, for~$\xi\in \mathbb{R}^d$ and~$x\in\mathcal{X}$, by
\[P_{A,B}(x)(\xi)=\frac{B(x)\cdot \xi}{A(x)\cdot B(x)}A(x).\]
For notational convenience, we introduce
\begin{equation}
    \label{eq:norton_covariation_projector_term}
    \Pi_{F,\nabla R,\sigma}(y)=\left[\overline{P}_{F,\nabla R} \sigma\sigma ^\intercal \overline{P}_{ F,\nabla R}^\intercal\right](y).
\end{equation}
We next proceed to identifying the bounded-variation increments on both sides of~\eqref{eq:norton_ito_formula}. After rearrangement and substitution of~\eqref{eq:norton_covariation_increment}, one obtains the following expression:

\begin{equation}
    \label{eq:norton_lambda_expr}
    \lambda_t^r =\lambda(Y_t^r),\qquad\lambda = -\frac1{F\cdot \nabla R}\left(b\cdot \nabla R +\frac12 \nabla^2 R:\Pi_{F,\nabla R,\sigma}\right).
\end{equation}
Substituting the expression for~$\Lambdar_t $ in~\eqref{eq:norton_dynamics} yields the following expression for the Norton dynamics:
\begin{equation}
    \label{eq:norton_dynamics_solved}
    \d \Yr_t = \overline{P}_{F,\nabla R}(\Yr_t)\left[b(\Yr_t) \d t + \sigma(\Yr_t)\, \d W_t\right]
    -\frac{\left(\nabla^2 R:\Pi_{F,\nabla R,\sigma}\right)(\Yr_t)}{2\nabla R(\Yr_t)\cdot F(\Yr_t)}F(\Yr_t)\,\d t.
\end{equation}
It can now be checked a posteriori, using Itô's formula, that the dynamics~\eqref{eq:norton_dynamics_solved} is such that~$R(\Yr_t)=r$ for all~$t\geq 0$, provided the coefficients are smooth. In Figure~\ref{fig:projector}, we illustrate geometrically the action of the projector~$\overline{P}_{F,\nabla R}$ at a point~$y$ on a vector~$\Delta y$.

Without further specifying the particular choice for the reference dynamics, the response flux observable~$R$ and non-equilibrium forcing~$F$, it is difficult to make general comments about the well-posedness of~\eqref{eq:norton_dynamics_solved}.
Let us however emphasize that a crucial condition for the dynamics to be well-defined is that the denominator~$F(\Yr_t) \cdot \nabla R(\Yr_t)$ in the expression of the projector~$P_{F,\nabla R}$ and in the last term of~\eqref{eq:norton_dynamics_solved} (which represents some curvature correction) should not vanish.
Thinking of the extreme case where~$\nabla R$ and~$F$ are everywhere orthogonal, we see that this requirement translates into a controllability condition: in this case, any forcing in the direction~$F$ has no effect on the flux, thus there is no way to control the latter using such a perturbation. More generally, starting from a configuration for which~$F\cdot \nabla R(q)=0$, it is not possible to maintain the value of the response function using the forcing~$F$. We therefore assume in the sequel that the condition
\begin{equation}
    \label{eq:norton_controllability_condition}
    \forall y\in\Sigma_r,\qquad F(y)\cdot \nabla R(y) \neq 0,
\end{equation}
is satisfied. We discuss the validity of this condition for the numerical examples we consider in Section~\ref{sec:numerical}.

\subsection{Norton analogs of the transport coefficient}\label{subsec:norton_tc}
We assume the well-posedness of the dynamics~\eqref{eq:norton_dynamics} (or equivalently~\eqref{eq:norton_dynamics_solved}), and also the existence and uniqueness of the invariant steady-state probability measure for this dynamics, whose expectation is denoted by~$\mathbb{E}_{r}^*$. We assume that
\begin{equation}
   \label{eq:lambda_centered_Norton}
   \mathbb{E}_0^*[\lambda]=0.
\end{equation}
This condition is the Norton counterpart of the centering  condition~\eqref{eq:R_centered_Thevenin} for the observable in usual NEMD simulations. When~\eqref{eq:lambda_centered_Norton} holds, the transport coefficient for the Norton dynamics is defined by analogy with~\eqref{eq:def_thevenin_transport_coeff} as
\begin{equation}
    \label{eq:def_norton_transport_coeff}
    \ct_{F,R}^* = \underset{r\to 0}{\lim}\,\frac{r}{\displaystyle{\mathbb{E}_r^* \left[\lambda\right]}},
\end{equation}
provided the limit exists. In equation~\eqref{eq:def_norton_transport_coeff},~$\ct_{F,R}^*$ can be interpreted as the inverse of the resistance to the non-equilibrium forcing. Note that the average forcing in the denominator of the right-hand side of~\eqref{eq:def_norton_transport_coeff} only involves the bounded varation part of~\eqref{eq:norton_lambda_sde}, since the expectation of the martingale part vanishes.

Provided the Norton dynamics is ergodic with respect to the steady-state, a natural estimator of the transport coefficient can be constructed by replacing ensemble averages by trajectory averages, similarly to what is done in standard NEMD simulations. More precisely, the estimator is computed using ergodic averages of~$\lambda(\Yr_t)$ under~$\E_r^*$,
that is,
\begin{equation}
\label{eq:norton_rho_estimator_continuous_time}
    \widehat{\ct}_{T,r}^*= \frac{rT}{\displaystyle{\int_{0}^T \lambda(Y_t^r)\,\d t}}.
\end{equation}

The statistical properties of this estimator can be analyzed similarly to what is done for the estimator~\eqref{eq:thevenin_rho_estimators_continuous_time}, leading to a result similar to~\eqref{eq:nemd_av_asymptotic}.
Assuming that a central limit theorem holds, and that~$F$,~$R$ are such that~$\E_0^*[\lambda]=0$, we can formally write the asymptotic variance associated with the estimator~\eqref{eq:norton_rho_estimator_continuous_time} as
\begin{equation}
    \label{eq:norton_av}
    \begin{aligned}
    \sigma_{\ct,r}^{2,*} &= \lim_{T\to \infty} T\Var^*_r(\widehat{\ct}_{T,r}^*)
     = \frac{r^4}{\E_{r}^*[\lambda]^4}\lim_{T\to \infty} T\Var^*_r\left(\frac{1}{\widehat{\ct}_{T,r}^*}\right)\\
    & = \frac{r^4}{\E_{r}^*[\lambda]^4}\frac{2}{r^2}\Var_r^*(\lambda)\Theta_r^*(\lambda),
        \end{aligned}
\end{equation}
where~$\Var_r^*$ denotes the variance with respect to~$\E_r^*$, and the Norton correlation time is defined similarly to the NEMD case by 
\begin{equation}
    \label{eq:norton_correlation_time}
    \Theta_{r}^*(\lambda) = \int_{0}^\infty \frac{\displaystyle{\E_r^*[\lambda(Y_t^r)\lambda(Y_0^r)]}}{\displaystyle{\E_r^*[\lambda^2]}}\,\d t.
\end{equation}
The second equality in~\eqref{eq:norton_av} follows from the Delta method~\cite[Chapter 3]{vdv00} applied to the reciprocal of~$\widehat{\ct}_{T,r}^*$. Using the first-order expansion given by~\eqref{eq:def_norton_transport_coeff}, and assumptions similar to the ones leading to~\eqref{eq:nemd_av_asymptotic}, we may further write, for~$|r|$ small:
\begin{equation*}
    \begin{aligned}
            \sigma_{\ct,r}^{2,*} &= \frac{2\left(\ct_{F,R}^{*}\right)^4}{r^2}\Var_r^*(\lambda)\Theta_r^*(\lambda) + \mathrm{O}\left(\frac1r\right)\\
    &= \frac{2\left(\ct_{F,R}^{*}\right)^4}{r^2}\Var_0^*(\lambda)\Theta_0^*(\lambda) + \mathrm{O}\left(\frac1r\right),
    \end{aligned}
\end{equation*}

As in the NEMD setting, one has to verify the validity of each of the expansion in powers of~$|r|$ for each of the quantities considered.
However, whereas in the NEMD setting this may be done in some cases by using a perturbative expansion of the non-equilibrium measure in powers of~$\eta$ (as done in~\cite{ss23}, for instance), there is in the Norton setting an additional technical difficulty in that the invariant measure for the Norton dynamics~\eqref{eq:norton_dynamics} is supported on the~$(d-1)$-dimensional manifold~$\Sigma_r$, which is disjoint from the Norton equilibrium manifold~$\Sigma_0$, and also of zero Lebesgue measure, so that it is both singular with respect to the invariant measure of the Norton equilibrium dynamics and to that of the reference dynamics~\eqref{eq:thevenin_dynamics}. A dedicated analysis has therefore to be performed to justify these approximations, and rigorously establish the validity of (non-)linear response formulas.

\subsection{Two straightforward generalizations}\label{subsec:norton_generalization}
    For ease of presentation, we restricted ourselves in Sections~\ref{subsec:norton_presentation} to~\ref{subsec:norton_tc} to the case where only one flux is fixed. However, the derivation of Norton dynamics can be straightforwardly extended to the case where several fluxes are simultaneously constrained. In fact, one can even consider situations where the response depends on time. These two generalizations can of course be combined.

    \paragraph{Multiple constraints.}
    We consider in this paragraph the case where~$c\geq 1$ forces $F_1,\dots,F_c$ act on the reference system with magnitudes~$\Lambda_{1,t}^{\mathbf{r}},\dots,\Lambda_{c,t}^{\mathbf{r}}$, chosen to maintain constant the value of~$c$ fluxes~$R_1,\dots,R_c$. The dynamics is given by
    \[\left\{\begin{aligned}\d Y_t^{\mathbf r} &= b(Y_t^{\mathbf r})\,\d t + \sigma(Y_t^{\mathbf r})\,\d W_t + \sum_{i=1}^c F(Y_t^{\mathbf r})\,\d \Lambda_i^{\mathbf r},\\ R_i(Y_t^{\mathbf r}) &= r_i,\qquad 1\leq i \leq c,\end{aligned}\right.\]
    where~$r_i$ gives the prescribed value of the corresponding flux~$R_i$.
    By defining the following maps defined on~$\mathcal X$
    \[R = \begin{pmatrix}R_1\\\vdots\\R_c\end{pmatrix} \in\mathbb{R}^c,\qquad F = \begin{pmatrix}|&  & |\\ F_1 & \dots & F_c \\ | &  & |\end{pmatrix}\in \mathbb{R}^{d\times c},\]
    as well as the following vectors,
    \[\Lambda_t^{\mathbf r}=\begin{pmatrix}\Lambda_{1,t}^{\mathbf r} \\ \vdots \\ \Lambda_{c,t}^{\mathbf r}\end{pmatrix}\in \mathbb{R}^{c},\qquad \mathbf{r}=\begin{pmatrix} r_1 \\ \vdots \\ r_c\end{pmatrix}\in\mathbb{R}^c,\]
    the Norton dynamics can be written similarly to~\eqref{eq:norton_dynamics}, upon replacing~$r$ by~$\mathbf{r}$:
    \begin{equation}
    \label{eq:norton_dynamics_rvec}
    \left\{\begin{aligned}
    \d Y^{\mathbf r}_t &= b(Y^{\mathbf r}_t)\, \d t + \sigma(Y^{\mathbf r}_t)\, \d W_t +  F(Y^{\mathbf r}_t)\,\d\Lambda^{\mathbf r}_t,\\
    R(Y^{\mathbf r}_t) &= R(Y^{\mathbf r}_0)=\mathbf{r}.
    \end{aligned}\right.
\end{equation}
    Computations identical to the ones leading to~\eqref{eq:norton_dynamics_solved} may be performed. As these are verbatim the same, we simply state the result and refer the reader to Appendix~\ref{appendix:multiple_constraints} for the complete derivation. The dynamics~\eqref{eq:norton_dynamics_rvec} can be written in closed form, using non-orthogonal projectors, as
    \begin{equation}
    \label{eq:norton_multiple_constraints}
        \begin{aligned}
        \d \Ybr_t &=\overline{P}_{F,\nabla R}(\Ybr_t)\left[b(\Ybr_t)\,\d t +\sigma(\Ybr_t)\,\d W_t\right]\\
        &\qquad- \frac{F(\Ybr_t)}2(\nabla R(\Ybr_t)^\intercal F(\Ybr_t))^{-1}\left(\nabla^2 R(\Ybr_t)^\intercal: \Pi_{F,\nabla R,\sigma}(\Ybr_t)\right),
    \end{aligned}
    \end{equation}
    where~$\nabla R : \mathcal X\to \mathbb{R}^{d\times c}$ and~$\nabla^2 R: \mathcal X \to \mathbb{R}^{d\times d\times c}$ are respectively the Jacobian and Hessian matrices of the fluxes. Here, we define the contraction product by
    \[\forall\, (A,B)\in \mathbb{R}^{d\times d\times c}\times\mathbb{R}^{d\times d},\qquad A:B = \left(\sum_{j,k=1}^d A_{kji}B_{kj}\right)_{1\leq i\leq c}\in\mathbb{R}^c,\]
    the projector~$\overline{P}_{F,\nabla R}$ is given, for~$A,B\in \mathbb{R}^{d\times c}$, by
    \[P_{A,B} = A(B^\intercal A)^{-1}B^\intercal,\qquad \overline{P}_{A,B} = \operatorname{Id}-P_{A,B},\]
    and~$\Pi_{F,\nabla R,\sigma}$ is defined similarly to~\eqref{eq:norton_covariation_projector_term} by
    \begin{equation*}
            \Pi_{F,\nabla R,\sigma}=\overline{P}_{F,\nabla R} \sigma\sigma ^\intercal \overline{P}_{F,\nabla R}^\intercal.
    \end{equation*}
    Note that the invariant measure is supported on a codimension~$c$ submanifold~$\Sigma_{\mathbf{r}}\subseteq \mathcal{X}$, and that the controllability condition becomes
    \begin{equation}
        \forall y\in \Sigma_{\mathbf{r}},\qquad\operatorname{det}(\nabla R(y)^\intercal F(y)) \neq 0.
    \end{equation}
    The average value of the forcing can be written as the ergodic average of the following vector-valued observable, which is the analog of~\eqref{eq:norton_lambda_expr}:
    \begin{equation}
        \lambda = -(\nabla R^\intercal F)^{-1}\left[\nabla R^\intercal b + \frac12\left(\nabla^2 R^\intercal: \Pi_{F,\nabla R,\sigma}\right)\right].
    \end{equation}
    Dynamics such as~\eqref{eq:norton_dynamics_rvec} should in particular allow for the numerical computation of Onsager relations in the Norton ensemble.
    
    \paragraph{Time-dependent fluxes.}
    One can also extend the dynamics to the case where we replace the condition~$R(Y_t^r)=r$ by~$R(Y_t^{\mathcal R})=\mathcal{R}_t$, when~$\mathcal{R}_t$ is the Itô process defined by
    \begin{equation}  
    \label{eq:norton_sde_R}
    \mathcal{R}_t=R(Y^{\mathcal R}_0) +\int_0^t \bar{r}_s\,\d s + \int_0^t \widetilde{r}_s\,\d B_s,
    \end{equation}
    with~$B$ a one-dimensional Brownian motion independent of~$W$, and~$\bar{r},\widetilde{r}$ stochastic processes adapted to the natural filtration of~$B$.
    These dynamics in particular cover the deterministic case~$\widetilde{r}_t = 0$, so that one can for instance consider time-periodic fluxes
    \[\bar{r}_t = \sin(2\pi \omega t),\qquad \widetilde{r}_t =0,\]
    or a stochastic process~$\Rr$ whose ergodic properties are well-understood, such as an Ornstein--Uhlenbeck process centered at~$r$:
    \[d\Rr_t = \gamma(r-\Rr_t)\,\d t +s\,\d B_t.\]
    One expects the resulting steady-state to be non-singular with respect to the invariant measure of the reference dynamics \eqref{eq:thevenin_dynamics}, which may be of some use from a theoretical perspective.
    The analog of~\eqref{eq:norton_dynamics} is given by
    \begin{equation}
    \label{eq:norton_dynamics_rvar}
    \left\{\begin{aligned}
    \d Y^{\mathcal R}_t &= b(Y^{\mathcal R}_t)\, \d t + \sigma(Y_t^{\mathcal R})\, \d W_t +  F(Y_t^{\mathcal R})\,\d\Lambda^{\mathcal R}_t,\\
    R(Y^{\mathcal R}_t) &= R(Y^{\mathcal R}_0)=\mathcal{R}^r_t.
    \end{aligned}\right.
    \end{equation}

    Following the same strategy as in the constant response case, one can express the dynamics in closed form as 
        \begin{multline}
        \label{eq:norton_time_varying_constraint}
        \d \Yr_t = \overline{P}_{F,\nabla R}(Y^{\mathcal R}_t)\left[b(Y^{\mathcal R}_t)\,\d t + \sigma(Y^{\mathcal R}_t)\,\d W_t\right]
        +\frac{\bar{r}_t\,\d t+\widetilde{r}_t\,\d B_t}{\nabla R(Y^{\mathcal R}_t)\cdot F(Y^{\mathcal R}_t)}F(Y_t^{\mathcal R})\\
        -\frac{1}{2\nabla R(Y^{\mathcal R}_t)\cdot F(Y^{\mathcal R}_t)}\left(\nabla^2 R:\left[ \frac{F\otimes F}{(\nabla R\cdot F)^2}\widetilde{r}_t^2 +\Pi_{F,\nabla R,\sigma}\right]\right)(Y^{\mathcal R}_t)F(Y^{\mathcal R}_t)\,\d t,
    \end{multline}
    and the bounded-variation contribution to the forcing as
    \begin{equation*}
    \label{eq:norton_lambda_t_expr}
    \lambda^{\mathcal R}_t = \left[\frac1{\nabla R\cdot F}\left(\bar{r}_t-b\cdot \nabla R -\frac12 \nabla^2 R:\left[  \frac{F\otimes F}{(\nabla R\cdot F)^2} \widetilde{r}_t^2 +\Pi_{F,\nabla R,\sigma}\right]\right)\right](Y^{\mathcal R}_t),
    \end{equation*}
    which is still an Itô process adapted to the larger filtration~$\left(\sigma(B_s,W_s : 0\leq s\leq t)\right)_{t\geq 0}.$ We refer to the Appendix~\ref{appendix:time_dependent_constraints} for details of the computations.

\section{Mobility and shear viscosity computations for Langevin dynamics}\label{sec:non_eq}
So far, we have made very few assumptions about the type of reference dynamics, driving force or flux. We now turn to presenting a framework in which the computation of physically meaningful transport coefficients may be performed, namely that of non-equilibrium (kinetic or underdamped) Langevin dynamics. We first recall the NEMD method for underdamped Langevin dynamics in Section~\ref{subsec:langevin_nemd}, before specifying the expressions of the non-equilibrium forcings and response observables relevant for mobility and shear viscosity computations in Section~\ref{subsec:langevin_tc}. We finally derive the corresponding Norton dynamics in Section~\ref{subsec:langevin_norton}, and interpret it in terms of a principle of least constraint.

\subsection{Standard non-equilibrium Langevin dynamics}\label{subsec:langevin_nemd}
The non-equilibrium framework we consider is that of a perturbation of the Langevin dynamics by an external driving force~$F$. 
The state of the system is described by a vector~$q\in \mathcal D$, describing the position of atoms, and a vector of corresponding momenta~$p\in \R^{dN}$, where~$d$ is the physical dimension and~$N$ is the number of atoms.
The configurational domain~$\mathcal D$ is usually considered to be~$\mathbb{R}^{dN}$ or~$(L\mathbb{T})^{dN}$ for some box length~$L>0$. The evolution of the system is governed by the Hamiltonian \[H(q,p) = V(q) +\frac12 p\cdot M^{-1}p,\]
where~$V:\mathcal D \to \R$ is the potential energy function. In practice,~$V$ is determined empirically to give an approximation of the ground-state energy for the Schr\"odinger Hamiltonian corresponding to the Born--Oppenheimer description of the system.
The dynamics is given by the following stochastic differential equation:
\begin{equation}
    \label{eq:langevin_equation}
    \left\{\begin{aligned}
    \d q_t & = M^{-1}p_t\, \d t, \\
    \d p_t &= \left[ -\nabla V(q_t) + \eta F(q_t) \right]\d t -\gamma M^{-1}p_t\,\d t +\sigma\,\d W_t.
    \end{aligned}\right.
\end{equation}
The parameter~$\beta > 0$ is proportional to the inverse temperature,~$M$ is a positive-definite symmetric mass matrix (typically a diagonal matrix), and~$\gamma$ and~$\sigma$ are~$dN\times dN$ matrices satisfying the fluctuation-dissipation relation 
\[\sigma\sigma^\intercal=\frac{2\gamma}\beta.\]
The friction coefficient~$\gamma$ is thus a symmetric positive semi-definite matrix, and~$W$ is a standard~$dN$-dimensional Brownian motion. Typically,~$\gamma$ is taken to be either a constant or a positive diagonal matrix, so that one simply writes~$\sigma=\sqrt{2\gamma/\beta}$. The parameter~$\eta$ once again governs the magnitude of the non-equilibrium perturbation.
One can show that at equilibrium~($\eta = 0$), the Boltzmann--Gibbs distribution
\begin{equation}
    \label{eq:boltzmann_gibbs_measure}
    \mu(q,p)\,\d q\,\d p= Z^{-1}\mathrm{e}^{-\beta H(q,p)}\,\d q\,\d p
\end{equation}
is an invariant probability measure for \eqref{eq:langevin_equation}.
In view of the separability of the Hamiltonian into a configurational and a kinetic part, the Boltzmann--Gibbs measure can be written in tensor form, as the product of a configurational measure
\[\mathit{m}(q)\,\d q = \frac1{Z_m}\mathrm{e}^{-\beta V(q)}\,\d q,\]
and of a Gaussian kinetic measure
\[\kappa(p)\,\d p  = \operatorname{det}\left(\frac{2\pi M}\beta\right)^{-1/2}\mathrm{e}^{-\frac\beta2 p\cdot M^{-1}p}\,\d p.\]
Momenta and positions are thus uncorrelated at equilibrium.
Since the perturbation is generally non-gradient, there is however no way to write out the expression of the invariant probability measure of the non-equilibrium dynamics ($\eta\neq 0$). Momenta and positions typically have a correlation of order~$\eta$ under this measure.

\subsection{Non-equilibrium forcings and fluxes}
\label{subsec:langevin_tc}
In this work, we consider transport coefficients corresponding to fluxes which can be written under the form
\begin{equation}
    \label{eq:nemd_response_observable}
    R(q,p)=G(q)\cdot p,
\end{equation}
where~$G: \mathcal D \to \mathbb{R}^{dN}$ is a vector field. 
Such forms of the response can be understood as measuring correlations between the momenta and some possibly non-linear feature of the configurational coordinates. In particular, such responses have zero-mean at equilibrium, owing to the tensor form of the Boltzmann--Gibbs measure and the fact that the average momentum under~$\kappa$ vanishes. The form~\eqref{eq:nemd_response_observable} is general enough to capture the cases of mobility and shear viscosity computations for molecular fluids, which we now proceed to present, and which are our numerical test cases.

\paragraph{Mobility computations.}
We first describe the NEMD method for the computation of diffusion properties.  Here, we consider a periodic domain~$\mathcal D = (L\mathbb{T})^d$.
The NEMD method is obtained by taking~$F$ as a constant vector field in \eqref{eq:langevin_equation}, and the velocity in the direction~$F$ as the response, which measures the particle flux in the direction~$F$. We assume that~$F$ is normalized as~$\| F \|=1$. Thus the perturbation and response observable are defined respectively by 
\begin{equation}
    \label{eq:mobility_force_flux}
    F(q) = F\in \mathbb{R}^{3N},\qquad R(q)= F\cdot M^{-1}p.
\end{equation}
Using the symmetry of~$M$, one may rewrite~$R(q,p)$ under the form~\eqref{eq:nemd_response_observable}. For practical computations, we consider two cases:
\begin{itemize}
    \item \emph{Single drift}: this corresponds to a perturbation where the force acts on a single component of the momentum, which can be assumed (by indistinguishability of the particles) to be the~$x$ component of the first particle: \begin{equation}\label{eq:single_drift}F_{\mathrm{S}}=(1,0,\dots)^\intercal \in \R^{dN}.\end{equation}
    \item \emph{Color drift} (see~\cite[Chapter 6]{em08}): this corresponds to a perturbation in which the force acts on half of the particles in one direction, and on half of the particles in the opposite direction, which we choose by convention to be the~$x$ direction: \begin{equation}\label{eq:color_drift}F_{\mathrm{C}}=\frac1{\sqrt N}(\underbrace{1,0,\dots,0}_{\in\, \R^d},-1,0,\dots,1,0\dots)^\intercal \in \R^{dN}.\end{equation}
\end{itemize}
The corresponding transport coefficients are related to the diffusion properties of the molecular system. More precisely, the transport coefficient~$\alpha_{F_{\mathrm{S}}}$ for the single drift forcing can be related to the diffusion coefficient
\[D = \underset{T\to\infty}{\lim}\,\frac1{2dNT}\sum_{i=1}^{dN}\E_0\left[ \left(\int_0^T M^{-1}p_{s,i}\,\d s\right)^2\right]\]for an isotropic system via
\begin{equation}
    \label{eq:diffusion_coefficient}
    \ct_{F_{\mathrm S}} = \beta D,
\end{equation}
as shown in~\cite{rodenhausen}.
When the potential energy function satisfies
\begin{equation}
    \label{eq:pairwise_potential_type_condition}
    \sum_{i=1}^N \frac{\partial}{\partial q_{i,x}}V(q)=0,
\end{equation}
a condition satisfied for instance for pairwise interactions depending only on the relative distances (as for~\eqref{eq:lennard_jones} below), then the transport coefficient~$\ct_{F_{\mathrm C}}$ for the color drift is related to~$\ct_{F_{\mathrm S}}$ through
\begin{equation}
\label{eq:color_drift_relation}
    \ct_{F_{\mathrm{C}}}=\ct_{F_{\mathrm S}}-\frac{2\lfloor N/2 \rfloor}{N(N-1)}\left(\frac1{\gamma}-\ct_{F_{\mathrm S}}\right).
\end{equation} 
See Appendix~\ref{appendix:color_drift} for the proof of this relation. In particular~$\ct_{F_{\mathrm{C}}}$ coincides with~$\ct_{F_{\mathrm{S}}}$ in the thermodynamic limit~$N\to\infty$.

The interest of considering these two drift perturbations is to assess whether the Norton method requires the forcing to act on the bulk of the sytem (color drift case) to be consistent with the NEMD method. The other extreme is that of a forcing acting on a single particle (single drift case).

\paragraph{Shear viscosity computations.}
The Norton framework described in this section also allows for the computation of the shear viscosity; see~\cite{TD07} for a review of various approaches to compute this transport coefficient. Instead of considering bulk forcings which require modifying the periodic boundary conditions (as studied in~\cite{Dobson} for instance), we consider the method introduced in~\cite{gms73}, for which the non-gradient force is periodic. The mathematical properties of this approach in the case of Langevin dynamics are studied in~\cite{js12}.
In this setting, we consider an anisotropic three-dimensional configurational domain of the form
\begin{equation}\label{eq:anisotropic_domain}\mathcal D = \left(L_x \mathbb{T} \times L_y\mathbb{T}\times L_z\mathbb{T}\right)^N.\end{equation}
We further allow for an anisotropic friction coefficient $\gamma$ which is diagonal and defined by three directional friction coefficients $\gamma_x, \gamma_y,\gamma_z\,> 0$.
A forcing is applied on the momenta in the longitudinal direction~$x$, with an intensity depending on the transverse configurational coordinate~$y$, according to a predefined forcing profile~$f:~L_y\mathbb{T}~\to~\mathbb{R}$:
\begin{equation}
\label{eq:shear_viscosity_forcing}
\forall\, 1\leq j\leq N,\qquad F(q)_{j,x}=f(q_{j,y}),\qquad F(q)_{j,y}=F(q)_{j,z} = 0.\end{equation}
In the non-equilibrium steady-state, the system displays an average longitudinal velocity profile depending only the transverse coordinate and the magnitude of the non-equilibrium perturbation.
More precisely, given an approximation of the identity~$\left(\psi_{\varepsilon}\right)_{\varepsilon>0}$ on~$L_y\mathbb T$, define
\begin{equation}
\label{eq:sv_response_profile}u_{x}(y)=\underset{\varepsilon\to 0}{\lim}\,\underset{\eta\to 0}{\lim}\, \frac{L_y}{\eta N}\,\E_\eta\left[\sum_{j=1}^N \left(M^{-1}p\right)_{j,x}\psi_{\varepsilon}(q_{j,y}-y)\right].\end{equation}
The term~$L_y$ in the numerator is motivated by the fact that the average with respect to the~$y$ coordinate of the term inside the limits is the velocity, up to a factor~$1/\eta$:
\[
\begin{aligned}  
\frac1{L_y}\int_{L_y\mathbb{T}} \left(\frac{L_y}{\eta N} \E_\eta\left[\sum_{j=1}^N \left(M^{-1}p\right)_{j,x} \psi_{\varepsilon}(q_{j,y} -y)\right]\right)\,\d y & = \frac1{\eta N}\sum_{j=1}^N \mathbb{E}_\eta\left[\left(M^{-1}p\right)_{j,x}\right] \\
& = \frac1{\eta} \mathbb{E}_\eta\left[\left(M^{-1}p\right)_{1,x}\right],
\end{aligned}
\]
using the indistinguishability of the particles to obtain the last equality.
Thus~$u_x$ corresponds to a localized linear response of the longitudinal velocity, which can be estimated in practice from trajectory averages using a binning procedure. One can then derive the following differential equation relating~$u_x$ to the shear viscosity~$\nu$:
\begin{equation}
  \label{eq:equation_viscosity}
  -\nu u_x''(y)+\gamma_x\rho u_x(y) = \rho f(y),
\end{equation}
where $\rho = N/(L_xL_yL_z)$ is the particle density of the system.
Since this profile is periodic in the transverse coordinate, the shear viscosity can be related to the Fourier coefficients of $u_x$ and $f$, namely
\[U_1=\frac{1}{L_y}\int_0^{L_y}u_x(y)\,\mathrm{e}^{\frac{2\mathrm{i}\pi y}{L_y}}\,\d y,\qquad F_1 = \frac{1}{L_y}\int_0^{L_y}f(y)\,\mathrm{e}^{\frac{2\mathrm{i}\pi y}{L_y}}\,\d y,\]
through
\begin{equation}
    \label{eq:shear_viscosity}
    \nu = \rho\left(\frac{F_1}{U_1}-\gamma_x \right)\left(\frac{L_y}{2\pi}\right)^2.
\end{equation}
For practical purposes, we choose~$f$ such that $F_1$ is analytically known, which is for instance the case if~$f$ is a sinusoidal profile.
Taking the limit~$\varepsilon\to 0$, in~\eqref{eq:sv_response_profile}, the Fourier coefficient~$U_1$ can be rewritten as
\begin{equation}
    U_1 = \lim_{\eta \to 0} \frac1{\eta N}\mathbb{E}_\eta \left[\sum_{j=1}^N\left(M^{-1}p\right)_{j,x}\mathrm{exp}\left(\frac{2 \mathrm{i}\pi q_{j,y}}{L_y}\right)\right].
\end{equation}
This expression is precisely the linear response~\eqref{eq:def_thevenin_transport_coeff} for the following response observable, which is an empirical Fourier coefficient for the longitudinal velocity profile:
\begin{equation}
    \label{eq:shear_response}
    R(q,p)=\frac{1}{N}\sum_{j=1}^N\left(M^{-1}p\right)_{j,x}\mathrm{exp}\left(\frac{2\mathrm{i}\pi q_{j,y}}{L_y}\right).
\end{equation}
This response is also of the form~\eqref{eq:nemd_response_observable}.
One can in turn estimate it using trajectory averages, as in~\eqref{eq:thevenin_rho_estimators_continuous_time}, yielding the estimator
\begin{equation}
    \label{eq:ux_estimator}
    \widehat{U}_{1,\eta,T} = \frac1{\eta T N}\sum_{j=1}^{N} \int_0^T \left(M^{-1} p_t\right)_{j,x}\exp\left(\frac{2\mathrm{i}\pi q_{j,y,t}}{L_y}\right)\d t.
\end{equation}

\subsection{The Norton method for Langevin dynamics}\label{subsec:langevin_norton}

We write more explicitly in this section the Norton dynamics for observables of the form~\eqref{eq:nemd_response_observable}. The general dynamics~\eqref{eq:norton_dynamics} greatly simplifies in this situation, and also admits a nice physical interpretation.

\paragraph{Closed form of the dynamics.}
Note first that~$\nabla^2_p R(q,p)=0$ when~$R$ is of the form~\eqref{eq:nemd_response_observable}, which together with the fact that the noise is degenerate (it acts only on the momenta in~\eqref{eq:langevin_equation}) implies that the quadratic covariation term in~\eqref{eq:norton_dynamics_solved} vanishes.
The forcing~$F$, too, acts only on the momenta, so that computing the projectors~\eqref{eq:norton_proj_AB} for the choices
\[A(q,p) = \begin{pmatrix}0\\ F(q)\end{pmatrix},\qquad B(q,p)=\nabla R(q,p) = \begin{pmatrix}\nabla G(q) p \\ G(q)\end{pmatrix},\]
yields the following expression for the projector associated with the Norton dynamics:
\[\mathrm{I}_{2dN}-\frac{A\otimes B}{A\cdot B}(q,p)=\begin{pmatrix}\displaystyle{\mathrm{I}_{dN}} & 0 \\ \displaystyle{-\frac{F(q)\otimes \nabla G(q)p}{F(q)\cdot G(q)}} & \displaystyle{\mathrm{I}_{dN}-\frac{F(q)\otimes G(q)}{F(q)\cdot G(q)}}\end{pmatrix}.\]
Since
\[\mathrm{I}_{dN}-\frac{F(q)\otimes G(q)}{F(q)\cdot G(q)}=\overline{P}_{F,G}(q),\]
where we slightly abuse the notation~\eqref{eq:norton_proj_AB}, it follows that the Norton dynamics~\eqref{eq:norton_dynamics_solved} is given by
\begin{equation}
    \label{eq:norton_langevin}
    \left\{
    \begin{aligned}
        \d q_t &= M^{-1}p_t\,\d t,\\
        \d p_t &= \overline{P}_{F,G}(q_t)\left(-\nabla V(q_t)\,\d t -\gamma M^{-1}p_t \,\d t +\sqrt{\frac{2\gamma}{\beta}}\,\d W_t\right) \\ &\quad-\frac{\nabla G(q_t)p_t\cdot M^{-1}p_t}{F(q_t)\cdot G(q_t)}F(q_t)\,\d t.
    \end{aligned}
    \right.
\end{equation}
Note that the configurational part of the dynamics is unaffected, which is consistent with the choice of a non-equilibrium perturbation acting solely on the momenta.
The forcing observable~\eqref{eq:norton_lambda_expr} is given by
\begin{equation}
    \label{eq:norton_lambda_langevin}
    \lambda(q,p) = \frac{G(q)\cdot(\nabla V(q) + \gamma M^{-1}p) -\nabla G(q) p \cdot M^{-1}p}{F(q)\cdot G(q)},
\end{equation}
and the controllability condition is given by
\begin{equation}
    \label{eq:norton_langevin_controllability}
    (G\cdot F)(q) \neq 0.
\end{equation}
\begin{remark}
We have restricted ourselves to a simple Langevin dynamics where the drift term is the sum of a gradient force and a linear friction term, and the noise is additive. The derivation can however be extended verbatim to a more general dynamics of the form
\begin{equation*}
    \left\{\begin{aligned}
        \d q_t&=M^{-1}p_t\,\d t,\\
        \d p_t&=b(q_t,p_t)\,\d t + \sigma(q_t,p_t)\,\d W_t,
    \end{aligned}\right.
\end{equation*}
for general drift terms~$b$ and diffusion matrices~$\sigma$.
For these kinetic dynamics, the associated Norton dynamics is given by
\begin{equation*}
    \left\{\begin{aligned}
        \d q_t&=M^{-1}p_t\,\d t,\\
        \d p_t&=\overline{P}_{F,G}(q_t)\left[b(q_t,p_t)\,\d t + \sigma(q_t,p_t)\,\d W_t\right]-\frac{\nabla G(q_t)p_t\cdot M^{-1}p_t}{F(q_t)\cdot G(q_t)}F(q_t)\,\d t.
    \end{aligned}\right.
\end{equation*}
\end{remark}

\paragraph{Physical interpretation.}
In the particular case of Langevin dynamics, the Norton method has the physical interpretation that it obeys a version of Gauss's principle of least constraint. The connection between Gauss's principle and non-equilibrium thermodynamics was already pointed out in~\cite{ehfml83}, see also the discussion in~\cite[Section~5.2]{em08}.
The principle of least constraints is a statement of classical dynamics equivalent to D'Alembert's principle, stating that the force applied to a system subject to a set of holonomic or non-holonomic constraints minimizes at every point in time the Euclidean distance to the force of the same system free from any constraints. Although we describe this interpretation in a deterministic setting, we stress that it remains valid, at least on a formal level,  upon considering a stochastic version of the dynamics.
More precisely, we assume that the dynamics for the unconstrained system can be written under the form
\begin{equation}
\label{eq:unconstrained_dynamics}
    \left\{
    \begin{aligned}
        \dot q &= M^{-1}p,\\
        \dot p &= f_{\mathrm{ref}}(q,p),
    \end{aligned}
    \right.
\end{equation}
and that the constraint is of the form~$R(q)=r$ (holonomic case) or~$R(q,p)=r$ (non-holonomic case). We can then write the dynamics of the constrained system as 
\begin{equation}
\label{eq:constrained_dynamics}
    \left\{
    \begin{aligned}
        \dot q &= M^{-1}p,\\
        \dot p &= f_{\mathrm{cons}}(q,p),
    \end{aligned}
    \right.
\end{equation}
where $f_{\mathrm{cons}}(q,p)$ is the force on the constrained system obeying Gauss's principle, which dictates that~$f_{\mathrm{cons}}(q,p)$ is the orthogonal projection of~$f_{\mathrm{ref}}(q)$ onto the affine hyperplane~$\mathcal{H}_{q,p}$ of admissible forces. This hyperplane can be determined by differentiating the constraint in time and setting it to zero. To obtain a constraint on the time derivative of the momenta~$\dot p$, this differentiation has to be performed once in the non-holonomic case and twice in the holonomic case. A simple computation shows that the hyperplane of admissible forces is thus given in the holonomic case by
\begin{equation}
\label{eq:gplc_hyperplanes_holo}
\mathcal H_{q,p}=\left\{\xi \in \R^{dN}\,\middle|\,\xi\cdot M^{-1}\nabla R(q) + \left(M^{-1}p\right)^{\otimes 2}:\nabla^2 R(q)=0\right\},
\end{equation}
and in the non-holonomic case by
\begin{equation}
    \label{eq:gplc_hyperplanes_nholo}
    \mathcal H_{q,p}=\left\{\xi\in \R^{dN}\,\middle|\, \xi\cdot \nabla_p R(q,p)+\left(M^{-1}p\right)\cdot \nabla_q R(q,p)=0\right\}.
\end{equation}
In particular, $f_{\mathrm{cons}}(q,p)-f_{\mathrm{ref}}(q,p)$ is orthogonal to~$\nabla_p R(q,p)$ in the non-holonomic case, and to~$M^{-1}\nabla R(q)$ in the holonomic case. The use of holonomic constraints is widespread in MD, where they are used to simulate systems with molecular constraints (such as fixed bond lengths or bond angles). See~\cite[Chapter 4]{lm15} for a general introduction, and~\cite{lrs12} for a detailed study of the underdamped Langevin case.

The Norton dynamics~\eqref{eq:norton_langevin} for a response function~$R$ of the form~\eqref{eq:nemd_response_observable} also satisfies a version of Gauss's principle of least constraint with respect to a non-Euclidean metric for which~$F(q)$ is everywhere orthogonal to~$G(q)^{\perp}$.
More precisely, this metric is induced by the configuration-dependent norm
\begin{equation}
    \label{eq:q_metric}
    \|\xi\|^2_q = \left(\xi\cdot F(q)\right)^2 + \left\|\xi-\frac{\xi\cdot G(q)}{\|G(q)\|^2}G(q)\right\|^2,
\end{equation}
where~$\|\cdot\|$ denotes the usual Euclidean norm. This metric is defined and non-degenerate if~$F$ and~$G$ are both non-zero, which is implied by the controllability condition~\eqref{eq:norton_controllability_condition}. As we show below, the force on the constrained system is, in this metric, obtained as
\begin{equation}
    \label{eq:variational_problem}
    \underset{\xi\in \mathcal{H}_{q,p}}{\argmin} \left\| \xi- f_{\mathrm{ref}}(q,p)\right\|_q = \overline{P}_{F,G}(q) f_{\mathrm{ref}}(q,p) - \frac{\nabla G(q) p\cdot M^{-1}p}{F(q)\cdot G(q)}F(q),
\end{equation}
where~$\mathcal{H}_{q,p}$ is the hyperplane defined in~\eqref{eq:gplc_hyperplanes_nholo} associated with the non-holonomic constraint~$R(q,p)=G(q)\cdot p=r$. This is precisely the Norton force corresponding to the unconstrained system~\eqref{eq:unconstrained_dynamics}.

To prove~\eqref{eq:variational_problem}, we proceed as follows. Note first that~$\mathcal{H}_{q,p}$ in~\eqref{eq:gplc_hyperplanes_nholo} is an affine translate of~$ \nabla_p R(q,p)^\perp = G(q)^\perp$, the Euclidean orthogonal to~$G(q)$, so that its normal direction with respect to the scalar product induced by~$\|\cdot\|_q$ is~$F(q)$.
Furthermore, the norm~$\|\cdot \|_q$ is constructed precisely so that the projector~$\overline{P}_{F,G}(q)$, whose action is depicted in Figure~\ref{fig:projector}, is an orthogonal projector onto~$G(q)^\perp$, for the scalar product associated with~$\|\cdot \|_q$. This implies that the minimizer in~\eqref{eq:variational_problem} is of the form
\[\xi^*(q,p) = f_{\mathrm{ref}}(q,p) -\alpha F(q) \in \mathcal{H}_{q,p}.\]
The value of~$\alpha$ is determined by the condition
\[\left(f_{\mathrm{ref}}(q,p) -\alpha F(q)\right)\cdot G(q) + \nabla G(q)p\cdot M^{-1}p =0,\]
hence 
\[\alpha = \frac{f_{\mathrm{ref}}(q)\cdot G(q) +\nabla G(q)p\cdot M^{-1}p}{F(q)\cdot G(q)},\]
and finally
\[\xi^*(q,p)=f_{\mathrm{ref}}(q)-\alpha F(q)=\overline P_{F,G}(q)f_{\mathrm{ref}}(q)-\frac{\nabla G(q)p\cdot M^{-1}p}{F(q)\cdot G(q)}F(q).\]
Loosely speaking, the Norton dynamics is the least non-equilibrium like of all dynamics on the constant response manifold, if one measures similarity in terms of the force considered in the~$\|\cdot\|_q$ metric.

\section{Numerical discretizations of Norton dynamics}\label{sec:discr}
We describe in this section a discretization of the Norton dynamics, first for the general dynamics~\eqref{eq:norton_dynamics} in Section~\ref{subsec:discr_gal}, before specializing it to the setting of Langevin dynamics. The numerical schemes are based for Langevin dynamics on splitting schemes, inspired by the ones typically used in the NEMD setting, but have the additional property of exactly preserving the flux throughout the numerical trajectory. For completeness, we describe splitting schemes in the case of standard NEMD Langevin dynamics in Section~\ref{subsec:discr_nemd}, and then in the Norton setting in Section~\ref{subsec:discr_norton}. We finally describe in Section~\ref{subsec:discr_lambda} a way to estimate the bounded variation part of the forcing~\eqref{eq:norton_lambda_expr} from the numerical trajectories, in order to approximate~\eqref{eq:norton_rho_estimator_continuous_time}.

\subsection{Numerical schemes for general Norton dynamics}\label{subsec:discr_gal}
We discuss here the simulation of the Norton dynamics~\eqref{eq:norton_dynamics}, which requires a discretization in time.
Formally, given a time step~$\Delta t >0$, a numerical scheme for the equilibrium dynamics~\eqref{eq:thevenin_dynamics} is defined by a map~$\Phi_{\Delta t} : \mathcal X \times \mathbb{R}^m \to \mathcal X$
which, iterated with independent and identically distributed (\iid) standard~$m$-dimensional Gaussian variables~$(\mathcal G_n)_{n\geq 0}$ yields a discretization of the dynamics:
\begin{equation}
    X^{n+1} = \Phi_{\Delta t}(X^n,\mathcal G^n),
\end{equation}
i.e. $X^n$ is an approximation of $X_{n\Delta t}$.
For example, the Euler--Maruyama scheme for the Th\'evenin dynamics~\eqref{eq:thevenin_dynamics} corresponds to
\begin{equation}
    \label{eq:em_scheme}
    \Phi^{\mathrm{EM}}_{\Delta t}(x,g)=x + \Delta t \left[b(x) +\eta F(x)\right] +\sqrt{\Delta t}\,\sigma(x)g.
\end{equation}
In principle, it would be possible to consider discretizations directly based on the autonomous form of the Norton dynamics~\eqref{eq:norton_dynamics_solved}, and average the forcing observable~$\lambda$ along the so-obtained numerical trajectories.
However, a key property we require from a numerical scheme is that the constant flux manifold should be preserved under the discrete dynamics. Standard discretizations of~\eqref{eq:norton_dynamics_solved} usually do not satisfy such preservation properties. 
Moreover, since the autonomous form of the dynamics involves second-order derivatives of the response, which may be cumbersome or expensive to compute, it is typically more convenient numerically to take another approach, obtained by enforcing the constraint via a Lagrange multiplier.
Given a stochastic scheme~$\Phi_{\Delta t}$ for the reference dynamics, we can consider the following discretization of the Norton dynamics:
\begin{equation}
    \label{eq:scheme_norton_general_sde}
    \left\{\begin{aligned}
        \widetilde{X}^{n+1} &= \Phi_{\Delta t}(X^n, G^n),\\
        X^{n+1} &= \widetilde{X}^{n+1} +\Delta t\Lambda^{n,*} F(X^n),\\
        R(X^{n+1}) &=r.
    \end{aligned}\right.
\end{equation}
Note that we chose here, somewhat arbitrarily, to perform the projection with respect to~$F(X^n)$. Other possible choices include~$F(\widetilde{X}^{n+1})$ or~$F(X^{n+1})$, the latter choice generally yielding an implicit scheme. In any case, finding~$\Lambda^{n,*}$ requires solving
\[R\left(\widetilde{X}^{n+1} +\Delta t\Lambda^{n,*} F(X^n)\right)=r,\]
which is typically a non-linear equation, for which the appropriate numerical strategy depends on the situation at hand. A typical choice is to resort to Newton's method or a fixed-point iteration, as done when enforcing holonomic constraints in MD (see~\cite[Section VII.1]{hlw06} and~\cite[Chapter~7]{lr04}), as well as~\cite{rcb77,a83} for pioneering works motivated by applications in MD.

Note that~$\Lambda^{n,*}$ corresponds to an approximation of the full forcing increment~$\Lambda_{(n+1)\Delta t}-\Lambda_{n\Delta t}$, and in particular incorporates the martingale increment. This martingale part should be removed when estimating~$\mathbb E_r^*[\lambda]$ in order to reduce the variance of the estimator under consideration. This can be done at first order in~$\sqrt{\Delta t}$ by using a control variate derived from its analytic expression~\eqref{eq:norton_forcing_martingale_part}, and the equality in law~$\sqrt{\Delta t} \, G^n \overset{\mathrm{law}}{=}W_{(n+1)\Delta t}-W_{n\Delta t}$. This leads to the following estimator of~$\lambda\Delta t$:
\begin{equation}
\label{eq:norton_var_reduction}
    \Lambda^{n} = \Lambda^{n,*}+\sqrt{\Dt}\frac{\nabla R(X^n)\cdot \sigma(X^n)G^n}{\nabla R(X^n)\cdot F(X^n)}.
\end{equation}
The efficiency of such variance reduction methods has already been demonstrated for the estimation of mean Lagrange multipliers associated with holonomic constraints, for example in the context of free energy computations (see~\cite{CLV08} and~\cite[Remark~3.33]{lrs10}).
An estimation of the ensemble average~$\mathbb E_r^*[\lambda]$ can be obtained by considering trajectory averages
\begin{equation}
    \label{eq:norton_lambda_n_definition}
    \widehat{\lambda}_{N_{\mathrm{iter}}} = \frac{1}{N_{\mathrm{iter}}} \sum_{n=1}^{N_{\mathrm{iter}}} \lambda^n,\qquad\lambda^{n} = \frac{1}{\Dt}\Lambda^n.
\end{equation}
As hinted at above, this strategy has the clear advantage that one does not have to compute~$\lambda(X^n)$ along the trajectory (which may be tedious to implement in practice), since consistent approximations thereof appear as natural byproducts of the integration procedure.

\subsection{Splitting schemes for (non-)equilibrium Langevin dynamics}\label{subsec:discr_nemd}
In the particular case where the reference dynamics is the Langevin dynamics~\eqref{eq:langevin_equation}, one can resort to a class of discretization strategies based on operator splittings of the infinitesimal generator, which we briefly recall for completeness in this section (see~\cite{lm13,lms16},  and~\cite[Chapter 7]{lm15} for more details).
By carefully choosing the order of the operators appearing in this splitting in the equilibrium setting, one can devise highly stable numerical schemes, which have a lower bias when sampling configurational averages in the overdamped limit~$\gamma\to\infty$ compared with other choices of the splitting order (see~\cite{lm13jcp}), while also showing desirable energy-conservation properties in the Hamiltonian limit~$\gamma\to 0$. Such schemes have become the standard choice to integrate kinetic Langevin dynamics in MD applications. The general procedure is as follows.

The generator of the (non-)equilibrium Langevin dynamics can be written as the sum of four terms,
\[\cL_{\eta} = \cL^{\mathrm{A}} + \cL^{\mathrm{B}} +\gamma\cL^{\mathrm{O}} +\eta \widetilde{\cL},\]
with 
\begin{equation}
    \label{eq:thevenin_generator_splitting_parts}
    \left\{
    \begin{aligned}
    \cL^{\mathrm{A}}&=M^{-1}p\cdot \nabla_q,\\
    \cL^{\mathrm{B}}&=-\nabla V\cdot \nabla_p,\\
    \cL^{\mathrm{O}}&=-M^{-1}p\cdot \nabla_p +\frac{1}{\beta}\Delta_p,\\
    \widetilde{\cL}&=F\cdot\nabla_p.
    \end{aligned}\right.
\end{equation}
The operators~$\cL^{\mathrm{A}}$,~$\cL^{\mathrm{B}}+\eta\widetilde{\cL}$ and~$\gamma\cL^{\mathrm{O}}$ can be viewed as infinitesimal generators in their own right, which correspond respectively to the three following elementary dynamics. The dynamics generated by~$\cL^{\mathrm{A}}$ is given by
\begin{equation}
\label{eq:thevenin_dynamics_splitting_A}
\left\{
\begin{aligned}
    \d q_t &= M^{-1}p_t\,\d t,\\
    \d p_t &= 0,
\end{aligned}\right.
\end{equation}
while~$\cL^{\mathrm{B}} +\eta\widetilde{\cL}$ gives rise to
\begin{equation}
\label{eq:thevenin_dynamics_splitting_B}
\left\{
\begin{aligned}
    \d q_t &= 0,\\
    \d p_t &= \left( -\nabla V(q_t) + \eta F(q_t)\right)\,\d t.
\end{aligned}\right.
\end{equation}
Finally,~$\cL^{\mathrm{O}}$ generates the following Ornstein--Uhlenbeck process on the momenta:
\begin{equation} 
    \label{eq:thevenin_dynamics_splitting_O}
    \left\{\begin{aligned}
    \d q_t &= 0,\\
    \d p_t &= -\gamma M^{-1}p_t \,\d t +\sqrt{\frac{2\gamma}{\beta}} \,\d W_t.
\end{aligned}\right.
\end{equation}
The three elementary dynamics~\eqref{eq:thevenin_dynamics_splitting_A},~\eqref{eq:thevenin_dynamics_splitting_B},~\eqref{eq:thevenin_dynamics_splitting_O} are analytically integrable. Splitting schemes for the Th\'evenin dynamics  are obtained by composing the evolution operators~$\e^{\Dt \cL^{\mathrm{A}}}$,~$\e^{\Dt \left(\cL^{\mathrm{B}} +\eta \widetilde{\cL}\right)}$ and~$\e^{\Dt \cL^{\mathrm{O}}}$ corresponding to each of these elementary dynamics. For instance, the evolution operator for the celebrated~$\mathrm{BAOAB}$ method corresponds to 
\[\e^{\frac{\Dt}2\left(\cL^{\mathrm{B}}+\eta\widetilde{\cL}\right)}\e^{\frac\Dt2\cL^{\mathrm{A}}}\e^{\Dt\gamma\cL^{\mathrm{O}}}\e^{\frac\Dt2\cL^{\mathrm{A}}}\e^{\frac{\Dt}2\left(\cL^{\mathrm{B}}+\eta\widetilde{\cL}\right)}.\]
These schemes can be formally justified, and rigorously analyzed with the Baker--Campbell--Hausdorff formula, yielding error estimates \`a la Talay--Tubaro on the invariant measure in the linear response regime where~$|\eta|$ is small, as well as on estimators of transport coefficients, see~\cite{lms16}.

\subsection{Splitting schemes for Langevin--Norton dynamics}\label{subsec:discr_norton}
For the Langevin dynamics~\eqref{eq:norton_langevin} in the Norton setting, a strategy similar to the one in Section~\ref{subsec:discr_nemd} can be used. The generator of the Norton dynamics can be split as the sum of three operators, each of which is the generator of an elementary dynamics, still denoted by~$\mathrm{A}$,~$\mathrm{B}$ and~$\mathrm{O}$. In fact, these are the Norton counterparts of the elementary dynamics~\eqref{eq:thevenin_dynamics_splitting_A},~\eqref{eq:thevenin_dynamics_splitting_B} and~\eqref{eq:thevenin_dynamics_splitting_O}, which therefore preserve the flux by construction. This splitting strategy is still motivated by the fact that each elementary dynamics individually preserves the flux observable~\eqref{eq:nemd_response_observable}.
However, in contrast to the NEMD case, the flow of the~$\mathrm{A}$-dynamics is generally not known in analytical form, and one has to resort to a numerical scheme to approximate it. 

In order to make the numerical scheme precise, the first step is to split the generator as
\begin{equation}
    \label{eq:norton_generator_splitting}
    \mathfrak{L} = \mathfrak{L}^{\mathrm{A}}+\mathfrak{L}^{\mathrm{B}}+\gamma\mathfrak{L}^{\mathrm{O}},
\end{equation}
with
\begin{equation}
    \label{eq:norton_generator_splitting_parts}
    \left\{
    \begin{aligned}
    \mathfrak{L}^{\mathrm{A}}&=M^{-1}p\cdot \nabla_q -\frac{\left(\nabla G\right) p \cdot M^{-1}p}{F\cdot G}F \cdot \nabla_p,\\
    \mathfrak{L}^{\mathrm{B}}&=-\overline{P}_{F,G}\nabla V\cdot \nabla_p,\\
    \mathfrak{L}^{\mathrm{O}}&=-\overline{P}_{F,G}M^{-1}p\cdot \nabla_p +\frac{1}{\beta} \overline{P}_{F,G}\overline{P}_{F,G}^\intercal:\nabla_p^2.
    \end{aligned}\right.
\end{equation}
This decomposition is motivated by the flux conservation properties~$\mathfrak{L}^{\mathrm{B}} R = \mathfrak{L}^{\mathrm{O}} R = 0$, which immediately imply that~$\mathfrak{L}^{\mathrm{A}}R=0$, since the overall dynamics with generator~$\mathfrak L$ conserves the non-equilibrium response by construction.

\paragraph{Decomposition into elementary dynamics.}
Analogously to the NEMD case, we interpret the various operators on the right-hand side of~\eqref{eq:norton_generator_splitting} as the generators of  elementary dynamics, and use the same terminology as in Section~\ref{subsec:discr_nemd}. In fact, these elementary dynamics are precisely the Norton counterparts~\eqref{eq:norton_dynamics_solved} to each of the NEMD elementary dynamics, so that they indeed individually preserve the response. They read
\begin{equation}\label{eq:norton_elem_dynamics}
   \begin{aligned}
        \text{$\mathrm{A}$ dynamics: }\qquad&\left\{\begin{aligned}
            \d q_t &= M^{-1}p_t \,\d t,\\
            \d p_t &= -\frac{\nabla G(q_t)p_t\cdot M^{-1}p_t}{F(q_t)\cdot G(q_t)}F(q_t)\,\d t.
        \end{aligned}\right.\\
        \text{$\mathrm{B}$ dynamics: }\qquad&\left\{\begin{aligned}
            \d q_t &= 0,\\
            \d p_t &=-\overline{P}_{F,G}(q_t)\nabla V(q_t)\,\d t.
        \end{aligned}\right.\\
        \text{$\mathrm{O}$ dynamics: }\qquad&\left\{\begin{aligned}
            \d q_t &= 0,\\
            \d p_t &= \overline{P}_{F,G}(q_t)\left(-\gamma M^{-1}p_t\, \d t +\sqrt{\frac{2\gamma}\beta}\,\d W_t\right).
        \end{aligned}\right.
    \end{aligned}    
\end{equation}
 Note that, in contrast to the NEMD setting, the~$\mathrm{A}$-dynamics is not analytically solvable, so that one has to resort to a numerical approximation similar to those discussed in Section~\ref{subsec:discr_gal} to integrate it (see Section~\ref{subsec:discr_lambda}). However, the Norton counterparts of the~$\mathrm B$ and~$\mathrm O$ dynamics remain analytically integrable, as we discuss below.
\paragraph{Analytic integration of the~$\mathrm B$ and~$\mathrm O$ dynamics.}
The~$\mathrm{B}$-dynamics is a time-linear evolution in the momenta, whose solution is given by
\[(q_t,p_t)=\left(q_0, p_0 - t \overline{P}_{F,G}\nabla V(q_0)\right)=\left(q_0,P_{F,G}(q_0)p_0+\overline{P}_{F,G}(q_0)\left[p_0-t\nabla V(q_0)\right]\right).\]
In view of the equality~$G(q)\cdot p = G(q)\cdot {P}_{F,G}(q) p$, and since~$\overline{P}_{F,G}$ is a projector onto~$G^\perp$, it is immediate that the~$\mathrm{B}$-dynamics preserves the response flux. One could also simply notice that it is the Norton counterpart to the NEMD~$\mathrm{B}$-dynamics~\eqref{eq:thevenin_dynamics_splitting_B}.

The~$\mathrm{O}$-dynamics is a projected Ornstein--Uhlenbeck process. To analytically integrate this dynamics, we assume that~$\overline{P}_{F,G}$ and~$\gamma M^{-1}$ commute. If this is not the case, the analytical integration below should be replaced by an appropriate numerical scheme, such as a midpoint rule. Using standard arguments of It\^o calculus, and the fact that~$\overline{P}_{F,G}$ is a projector, it is straightforward to check that the solution is given for all~$t\geq 0$ by~$q_t=q_0$ and 
\[p_t = {P}_{F,G}(q_0)p_0 +\overline{P}_{F,G}(q_0)\left(\e^{-t\gamma M^{-1}}p_0 +\int_0^t \e^{-\gamma M^{-1}(t-s)}\sqrt{\frac{2\gamma}{\beta}}\,\d W_s\right).\]
This is a Gaussian process, which has the following alternative representation  in distribution:
\begin{equation}
\label{eq:norton_o_step_gaussian_distribution}
p_t = {P}_{F,G}(q_0)p_0 + \overline{P}_{F,G}(q_0)\left(\e^{-t\gamma M^{-1}}p_0 +\sqrt{\frac{1-\e^{-2t\gamma M^{-1}}}{\beta}M}\mathcal{G}\right),
\end{equation}
where~$\mathcal{G}$ is a standard~$dN$-dimensional Gaussian. By the same arguments as for the~$\mathrm{B}$-dynamics, the Norton~$\mathrm{O}$-dynamics preserves the response flux, and is likewise the Norton counterpart to the NEMD~$\mathrm{O}$-dynamics~\eqref{eq:thevenin_dynamics_splitting_O}.

\subsection{Estimation of the average forcing}\label{subsec:discr_lambda}
By composing the evolution operators of elementary Norton dynamics, one obtains a natural splitting approximation of the evolution operator for the forcing process of the full Norton dynamics. The numerical translation of this observation is that one can estimate trajectory averages of~$\lambda$ directly from examining individual integration steps of the splitting scheme. We describe in this section the general procedure to this end.

For a fixed time step~$\Delta t>0$ and response magnitude~$r\in\mathbb R$, we define three discrete flows acting on the augmented state~$(q,p,\ell)\in \mathcal D \times \R^{dN}\times \R$. The role of the auxiliary variable~$\ell$ is to accumulate the bounded-variation component of Lagrange multipliers enforcing the constant-flux constraint during the integration step. It is thus initialized at~$0$ at the start of each integration step, accumulating the contributions of each part of the splitting.

\paragraph{Discrete integration of the~$\mathrm A$ dynamics.}
The discrete flow associated with the~$\mathrm{A}$-step is given by
\begin{equation}
    \label{eq:phi_A}
    {\Phi_{\Delta t,r}^{\mathrm{A}}\left(q,p,\ell\right) = \left(q +\Delta t M^{-1}p,p+\xi_{\Delta t,r}^{\mathrm{A}}(q,p) F\left(q +\Delta t M^{-1}p\right),\ell+\xi^{\mathrm{A}}_{\Delta t,r}(q,p)\right)},\end{equation}
where~$\xi_{\Delta t,r}^{\mathrm{A}}(q,p)$ is the Lagrange multiplier characterized by the constant-flux condition
\[R\left(\Phi_{\Delta_t,r}^{\mathrm{A}}(q,p,\ell)\right) = r,\]
with some abuse of notation in the arguments of~$R$.
When the response is of the form~$R(q,p)= G(q)\cdot p$, we can explicitly solve for the Lagrange multiplier, as
\begin{equation}
    \label{eq:xi_A}
    \xi_{\Delta t,r}^{\mathrm{A}}(q,p)= \frac{r-G\left(q +\Delta t M^{-1}p\right)\cdot p}{F\left(q +\Delta t M^{-1}p\right)\cdot G\left(q +\Delta t M^{-1}p\right)}.
\end{equation}
The Norton~$\mathrm{A}$-step is therefore equivalent to first evolving the system's coordinate variables according to the corresponding NEMD~$\mathrm{A}$-step before correcting the momenta, by applying a constraining force in the direction of~$F(q+\Delta t M^{-1}p)$, the forcing evaluated at the updated position. As mentioned after~\eqref{eq:scheme_norton_general_sde}, there is some freedom in the choice of point at which to evaluate the forcing when projecting onto the submanifold~$\Sigma_r = R^{-1}\{r\}$. The motivation for choosing~$F(q+\Delta t M^{-1}p)$ is that the two functions appearing in the scalar product~$F\cdot G$ in the denominator of the left-hand side of~\eqref{eq:xi_A} are evaluated at the same point. 

\paragraph{Discrete integration of the~$\mathrm B$ dynamics.}
The discrete flow associated with the Norton~$\mathrm{B}$-step is given by
\begin{equation}
    \label{eq:phi_B}
    \Phi_{\Delta t,r}^{\mathrm{B}}(q,p,\ell) = \left(q, p - \Delta t\nabla V(q) + \xi_{\Delta t,r}^{\mathrm{B}}(q,p)F(q), \ell + \xi_{\Delta t,r}^{\mathrm{B}}(q,p)\right),
\end{equation}
where~$\xi_{\Delta t,r}^{\mathrm{B}}$ is again a Lagrange multiplier enforcing the constraint, given for a response~$G(q)\cdot p$ by
\begin{equation}
    \label{eq:xi_B}
    \xi_{\Delta t,r}^{\mathrm{B}}(q,p)= \frac{r-G\left(q\right)\cdot \left(p -\Delta t \nabla V(q)\right)}{F\left(q\right)\cdot G\left(q\right)}.
\end{equation}
This again corresponds to a step of the NEMD~$\mathrm{B}$-dynamics, re-projected onto~$\Sigma_r$ in the direction~$F(q)$. This coincides in fact with the analytic integration of the elementary~$\mathrm{B}$-dynamics~\eqref{eq:norton_elem_dynamics} when~$G(q)\cdot p = r$.

\paragraph{Discrete integration of the~$\mathrm O$ dynamics.}
The flow associated with the Norton~$\mathrm{O}$-step is stochastic, hence we formulate it conditionally on a standard~$dN$-dimensional Gaussian~$\mathcal{G}$ increment. In fact, we show below that a convenient update is
\begin{equation}
    \label{eq:phi_O}
    {
    \Phi_{\Delta t,r}^{\mathrm{O}}
        \left(q,p,\ell \middle|\, \mathcal G \right)=
        \left(
        q,\alpha_{\Delta t}p + \sigma_{\Delta t}\,\sqrt{\Delta t}\, \mathcal{G} + \xi_{\Delta t,r}^{\mathrm{O}}(q,p,\mathcal G) F(q),\ell + \frac{r(1-\alpha_{\Delta t})}{F(q)\cdot G(q)}
        \right)}
    ,
\end{equation}
where~$\alpha_{\Dt}$ and~$\sigma_{\Dt}$ are given by \[\alpha_{\Delta t} = \e^{-\gamma M^{-1}\Dt},\,\qquad \sigma_{\Delta t} = \sqrt{\frac{1-\alpha_{\Dt}^2}{\beta\Delta t}M}.\]
To motivate~\eqref{eq:phi_O}, we start by deriving the expression of the Lagrange multiplier~$\xi_{\Delta t,r}^{\mathrm{O}}$, which can be solved for our particular form of response~\eqref{eq:nemd_response_observable} as
\begin{equation}
    \label{eq:xi_O}
    \xi_{\Dt,r}^{\mathrm{O}}(q,p,\mathcal G) = \frac{r - \alpha_{\Dt}G(q)\cdot p -\sigma_{\Dt}\sqrt{\Dt} \, G(q)\cdot {\mathcal G} }{F(q)\cdot G(q)}.
\end{equation}
Again, this corresponds to integrating the NEMD~$\mathrm{O}$-dynamics over one time step before correcting the momenta in the direction~$F(q)$. Since the contribution of~$\mathcal{G}$ to this~$\xi_{\Delta t,r}^{\mathrm O}$ is a centered Gaussian, we can remove it entirely to only accumulate the non-martingale component in~$\ell$. This is exactly equivalent to the variance reduction technique discussed in~\eqref{eq:norton_var_reduction}. Using~$G(q)\cdot p = r$, one arrives at the expression given in~\eqref{eq:phi_O} for the action of~$\Phi_{\Dt,r}^{\mathrm{O}}$ on~$\ell$. Once again,~$\Phi_{\Dt,r}^{\mathrm{O}}(q,p,\ell)_{q,p}$ corresponds to the analytic flow of the Norton~$\mathrm{O}$-dynamics~\eqref{eq:norton_elem_dynamics} over one time step~$\Delta t$, with deterministic initial condition~$(q,p)$.

\paragraph{Construction of the splitting scheme.}
The steps~$\mathrm{A,B,O}$ can be composed according to the order of the splitting, and the bounded-variation increment of the constraining process over one time step can be estimated from the final increment in~$\ell$. For example, conditionally on~$\mathcal{G}^{n+1}$, a standard~$dN$ dimensional Gaussian, the update rule for the Norton BAOAB scheme is given by
\begin{equation}\label{eq:baoab_oneline}{(q^{n+1},p^{n+1},\ell^{n+1}) = \Phi_{\Dt/2,r}^{\mathrm{B}}\circ \Phi_{\Dt/2,r}^{\mathrm{A}} \circ \Phi_{\Dt,r}^{\mathrm{O}}(\cdot\,|\,\mathcal G^{n+1})\circ \Phi_{\Dt/2,r}^{\mathrm{A}}\circ\Phi_{\Dt/2,r}^{\mathrm{B}} (q^n,p^n,0)}.\end{equation}
Note that we start from~$\ell^n = 0$, and then accumulate the various increments of the constraining process to obtain~$\ell^{n+1}$.
For concreteness, let us explicitly write the update rules for the numerical scheme~\eqref{eq:baoab_oneline} and a response observable~$R(q,p)=G(q)\cdot p$:
\begin{equation}
    \label{eq:detailed_baoab}
    \left\{\begin{aligned}
        p^{n+\frac15} &= p^{n}-\frac{\Delta t}2 \nabla V(q^n) + \xi_{\Delta t/2,r}^{\mathrm B}(q^n,p^n)F(q^n),\\
        \ell^{n+\frac15} &= \xi_{\Delta t/2,r}^{\mathrm B}(q^n,p^n),\\
        q^{n+\frac25} &= q^n +\frac{\Delta t}2 M^{-1}p^{n+\frac15},\\
        p^{n+\frac25} &= p^{n+\frac15}+\xi_{\Delta t/2,r}^{\mathrm A}(q^n,p^{n+\frac15})F(q^{n+\frac25}),\\
        \ell^{n+\frac25} &= \ell^{n+\frac15}+\xi_{\Delta t/2,r}^{\mathrm A}(q^n,p^{n+\frac15}),\\
        {p}^{n+\frac 35} &= \alpha_{\Delta t}p^{n+\frac25} + \sigma_{\Delta t} \mathcal{G}^{n+1}+\xi_{\Delta t,r}^{\mathrm O}(q^{n+\frac25},p^{n+\frac25},\mathcal{G}^{n+1})F(q^{n+\frac25}),\\
        \ell^{n+\frac35}&=\ell^{n+\frac25}+\frac{r(1-\alpha_{\Delta t})}{F(q^{n+\frac25})\cdot G(q^{n+\frac25})},\\
        q^{n+1} &= q^{n+\frac25} +\frac{\Delta t}2 M^{-1}p^{n+\frac35},\\
        p^{n+\frac45} &= p^{n+\frac35}+\xi_{\Delta t/2,r}^{\mathrm A}(q^{n+\frac25},p^{n+\frac35})F(q^{n+1}),\\
        \ell^{n+\frac45} &= \ell^{n+\frac35}+\xi_{\Delta t/2,r}^{\mathrm A}(q^{n+\frac25},p^{n+\frac35}),\\
         {p}^{n+1} &= p^{n+\frac45}-\frac{\Delta t}2 \nabla V(q^{n+1})+\xi_{\Delta t/2,r}^{\mathrm B}(q^{n+1},{p}^{n+\frac45})F(q^{n+1}),\\
        \ell^{n+1} &= \ell^{n+\frac45}+\xi_{\Delta t/2,r}^{\mathrm B}(q^{n+1},{p}^{n+\frac45}).
        \end{aligned}\right.
\end{equation}
The average value of~$\lambda$ over the corresponding time step can then be estimated via
\begin{equation}
\label{eq:norton_discrete_forcing_estimator}
    \lambda^{n+1} = \Dt^{-1}\ell^{n+1}.
\end{equation}

Since all the substeps are by construction preserving the value of the response function, splitting schemes based on~\eqref{eq:norton_generator_splitting} define discrete-time Markov chains on~$\Sigma_r$. Fixing a splitting and a time step~$\Delta t$, one can hope that the Markov chain admits a unique invariant probability measure~$\pi^*_{r,\Dt}$ close to the invariant measure of the continuous time Norton dynamics for~$\Delta t>0$ sufficiently small. Assessing the quality of these schemes (measured in terms of weak or strong error estimates, or errors on the invariant measure) as a function of the time step~$\Dt$, the magnitude of the perturbation~$r$ and the ordering of the splitting is an open question.

\begin{remark}
Let us emphasize that, when the form of the response function does not allow for analytical expressions of the Lagrange multipliers, one can still apply a splitting procedure similar to~\eqref{eq:detailed_baoab}, upon replacing the explicit expressions of~$\xi_{\Delta t,r}^X$ (for~$X \in \{\mathrm{A},\mathrm{B},\mathrm{O}\}$) by the numerical solution of some nonlinear equation determining these quantities. It is still possible to cancel at dominant order the martingale increment by subtracting the first order approximation in~$\sqrt{\Delta t}$ of the latter quantity. 
\end{remark}

\section{Numerical results}\label{sec:numerical}
We present in this section the results of various numerical simulations for the Lennard--Jones fluid.
The numerical experiments we perform have several aims:
\begin{itemize}
\item {The first one is to verify that, at least in our setting, the Norton dynamics is a viable approach for the computation of transport coefficients, in the sense that Norton estimators of the linear response coincide with those obtained from NEMD computations. Of course, this cannot be expected for low dimensional systems. Consider for example the case of  kinetic Langevin dynamics for a single one-dimensional particle on the unit torus~$\mathbb{T}$, with~$ M = F = 1$, and~$R(q,p) = p$. The invariant measure for the Norton dynamics is then easily seen to be the product of the uniform measure on 
$\mathbb{T}$ with the Dirac measure~$\delta_r$ on momentum space, from which it follows in view of~\eqref{eq:norton_lambda_langevin} that
\[\E_r^*[\lambda] = \int_0^1 V'(q)\,\d q +\gamma r = \gamma r,\]
since~$V$ is periodic, whence the Norton transport coefficient is~$\gamma^{-1}$, which differs in general from its NEMD counterpart.}

\item {Once the validity of the Norton approach has been established, the second aim is to assess the numerical efficiency of the Norton method, relatively to the standard NEMD method. A crucial point is to determine whether the Norton approach leads to a reduction in the asymptotic variance~\eqref{eq:norton_av} compared to~\eqref{eq:nemd_av_asymptotic} for estimators of the transport coefficient, as this quantity determines the magnitude of the statistical error, and therefore the simulation time needed to attain a given level of accuracy.}
\end{itemize}

In Section~\ref{subsec:numerical_system}, we present the system and notation used throughout all numerical experiments. In Section~\ref{subsec:numerical_consistency}, we show instances of excellent agreement in the linear response between Norton and NEMD dynamics. We also exhibit a case in which the response profiles do not coincide. We further show that the agreement (when it holds) extends far into the non-linear response regime. In Section~\ref{subsec:numerical_thermo_limit}, we investigate the properties of the Norton and NEMD systems in the thermodynamic limit, showing that the linear responses coincide far before convergence to the thermodynamic limit, and exhibiting an interesting concentration property for the distribution of values of~$\lambda$, both at equilibrium and in the non-equilibrium regime. Finally, in Section~\ref{subsec:numerical_av} we assess the efficiency of the Norton approach in terms of asymptotic variance, showing that in some cases the standard NEMD approach is significantly outperformed.

\subsection{Description of the numerical experiments}\label{subsec:numerical_system}
\paragraph{Presentation of the system.}
In all experiments, we consider perturbations of the kinetic Langevin dynamics for a Lennard--Jones fluid. The potential energy function is given by
\begin{equation}
        \label{eq:lennard_jones}
        V_{\mathrm{LJ}}(q)=\sum_{1\leq i < j \leq N} v\left(\|q_i-q_j\|\right),        
    \end{equation}
    where~$v$ is the radial function
   ~$$v(r)=4\varepsilon \left( \left( \frac{\sigma}{r}\right)^{12}-\left(\frac{\sigma}{r} \right)^6\right).$$
   Note that~$v'(2^{1/6}\sigma)=0$.
   The parameters~$\varepsilon$ and~$\sigma$ modulate respectively the energy and spatial range of the interaction. It is convenient to state results in the system of reduced units in which~$\varepsilon$,~$2^{1/6}\sigma$, the mass~$m$ of each particle and the Boltzmann constant~$k_{\mathrm{b}}$ are set to~$1$.  In fact, we consider a slightly modified version of the potential~\eqref{eq:lennard_jones}, obtained by truncating the range of~$v$ as
    \[v_{r_{\mathrm{c}}}(r)=\left[v(r)-v(r_{\mathrm c}) -v'(r_{\mathrm c})(r-r_{\mathrm c})\right]\mathbbm{1}_{r\leq r_{\mathrm{c}}},\]
   where~$r_{\mathrm c}$ is a cutoff radius which prescribes the maximum range of interactions, the added affine term ensuring that~$v_{r_{\mathrm{c}}}$ is~$C^1$.
   We set~$r_{\mathrm c} = 2.5$ in our simulations, which were performed using the Julia package {\tt Molly}~\cite{molly}. Both the Norton and NEMD simulations were carried out using a~$\mathrm{BAOAB}$ numerical splitting scheme.
   
    \paragraph{Discrete estimators of transport coefficients.}
    We compute approximations of the mobility by plotting the average response as a function of the forcing magnitude~$\eta$, and fitting the slope of the initial linear regime to obtain the transport coefficient. More precisely, continous time estimators for the transport coefficient~\eqref{eq:thevenin_rho_estimators_continuous_time} and its Norton analog~\eqref{eq:norton_rho_estimator_continuous_time} can be obtained by defining appropriate discretizations thereof. These discretizations are given by the following estimators
    \begin{equation}
    \label{eq:discr_estimators_ct}
    \widehat{\ct}^{\Delta t}_{N_{\mathrm{iter}},\eta} = \frac{1}{N_{\mathrm{iter}}\eta}\sum_{k=0}^{N_{\mathrm{iter}}-1}R(q^k,p^k), \qquad  \widehat{\ct}^{\Delta t,*}_{N_{\mathrm{iter}},r} = r N_{\mathrm{iter}}\left( \sum_{k=0}^{N_{\mathrm{iter}}-1} \lambda^k\right)^{-1},
    \end{equation}
    respectively in the NEMD case and the Norton case, where~$\lambda^k$ is defined in~\eqref{eq:norton_discrete_forcing_estimator} for the Norton dynamics preserving some response~$R$. More reliable estimators can further be obtained by fitting the response profile for several small values of~$\eta$ and~$r$.

    In the case of mobility computations, the observable~$R(q,p) = F\cdot M^{-1}p$ is used, yielding the following NEMD estimator for the mobility:
    \begin{equation}
        \widehat{\ct}^{\Delta t}_{N_{\mathrm{iter}},\eta} = \frac1{N_{\mathrm{iter}}\eta}\sum_{k=0}^{N_{\mathrm{iter}}-1} F\cdot M^{-1}p^k.
    \end{equation}
    For shear viscosity, discretizations of the NEMD estimator~\eqref{eq:ux_estimator} for the Fourier coefficient~$U_1$ are used, yielding
    \begin{equation}
        \widehat{U}_{1,N_{\mathrm{iter}},\eta}^{\Delta t} = \frac1{N N_{\mathrm{iter}}\eta}\sum_{k=0}^{N_{\mathrm{iter}}}\sum_{j=1}^{N} \left(M^{-1}p^k\right)_{j,x}\exp\left( \frac{2 \mathrm{i} \pi q^k_{j,y}}{L_y}\right).
    \end{equation}
    
    \paragraph{Verification of the controllability condition.}
    We conclude this section by discussing the controllability condition~\eqref{eq:norton_langevin_controllability} in the case of mobility and shear viscosity computations.

    When computing the mobility using a constant forcing, note that, using the expression for the non-equilibrium response in~\eqref{eq:mobility_force_flux}, the controllability condition writes
    \[(G\cdot F)(q) = F^\intercal M^{-1}F \neq 0,\]
    since~$M^{-1}$ is a positive definite matrix, so that the controllability condition is trivially satisfied everywhere. Moreover, the existence of strong solutions to~\eqref{eq:norton_langevin} can be straightforwardly deduced from the existence of strong solutions to~\eqref{eq:langevin_equation}, since in this case~$\overline{P}_{F,G}=\overline{P}_{F,M^{-1}F}$ is a constant matrix, which therefore preserves locally Lipschitz maps which grow at most linearly at infinity, see for instance the discussion in~\cite[Section 3.3]{PavliotisBook}.

    We examine the controllability condition in the case of a transverse shear forcing profile in the case~$M=m\mathrm{Id}$ in order to alleviate the notation (the extension to more general mass matrices being straightforward). Using the expression for the response~\eqref{eq:shear_response}, the controllability condition writes
    \[(G\cdot F)(q) = \frac{1}{mN}\sum_{j=1}^N \exp\left({\frac{2\mathrm{i}\pi q_{j,y}}{L_y}}\right)f(q_{j,y}) \neq 0.\]
    The quantity $G\cdot F$ could in principle vanish. As the number of particles is increased, this should however be rather unlikely when the first Fourier coefficient of the forcing~$f$ is non-zero. Indeed, the marginal distribution of a single particle is uniform over the simulation cell at equilibrium (by translation invariance). Therefore, one expects, for a large number of particles and not too strong a forcing, that $G\cdot F$ is close to
    \[
    \frac{1}{m L_y} \int_{0}^{L_y} \exp\left({\frac{2\mathrm{i}\pi y}{L_y}}\right)f(y) \, \d y,
    \]
    which is exactly the first Fourier coefficient of~$f$, a non-zero quantity.

    \begin{subsection}{Equivalence of (non-)equilibrium responses}\label{subsec:numerical_consistency}
    We begin by checking the consistency between the standard NEMD approach and the Norton approach, by applying these two methods to the cases of mobility and shear viscosity computations discussed in Section~\ref{subsec:langevin_tc}.
    \paragraph{Mobility computations.}
    We begin by checking the validity of the Norton method on the simple example of mobility. We used identical equilibrium conditions for NEMD and Norton computations, namely a three-dimensional system of~$N=1000$ particles, in a cubic periodic domain of side length~$L$, such that~$\rho=N/L^3 = 0.6$, with a temperature~$T = 1.25$ and a friction coefficient~$\gamma = 1.0$. The time step was set to~$\Dt = 10^{-3}$ (this choice ensures that the relative variations of the energy of the system are of the order of~0.6 for Hamiltonian dynamics), and simulations were performed for a minimal physical time of~$T_{\mathrm{sim}} = 250,000$ in the linear response regime, and~$T_{\mathrm{sim}} = 8,000$ far in the non-linear response regime. Error bars have been omitted for the sake of readability, as they are in all cases smaller than the size of the markers.

    In Figure~\ref{fig:mobility_linear}, we plot the response as a function of the forcing magnitude for both the Norton and NEMD dynamics. Thus, the fixed quantity is plotted in the ordinates for the Norton system, and in the abscissas for the NEMD system. Least-squares linear regression lines are plotted, in a dotted red line for the Norton system, and a dashed green line for the NEMD system. The slopes of these lines, indicated in the legend, give the estimated transport coefficient. The left plot corresponds to a color drift perturbation~\eqref{eq:color_drift}, while the right plot corresponds to a single drift perturbation~\eqref{eq:single_drift}. We observe that, while the agreement is almost perfect in the case of the color-drift forcing, there is a significant discrepancy in the linear responses in the case of a single-drift forcing. In fact, using the relation~\eqref{eq:color_drift_relation}, it is readily checked that the Norton estimation of the diffusion coefficient is incorrect. We believe that for the Norton method to be valid, the microscopic perturbation should as a general rule act on the bulk of the system. This condition is not satisfied in the case of a single-drift forcing, which acts on a single particle.
    
        \begin{figure}
        \centering
        \includegraphics[width=0.49\textwidth]{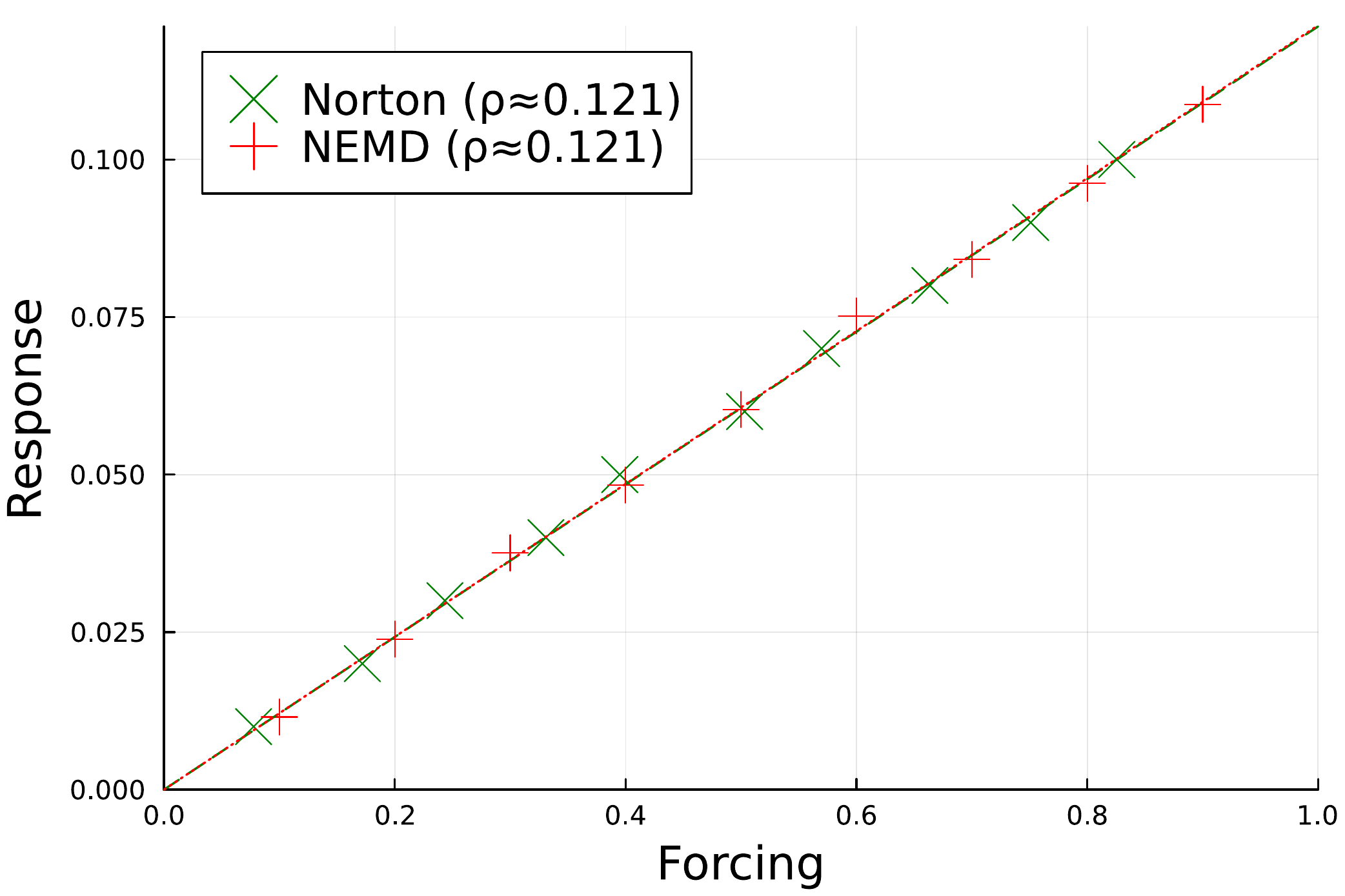}
        \includegraphics[width=0.49\textwidth]{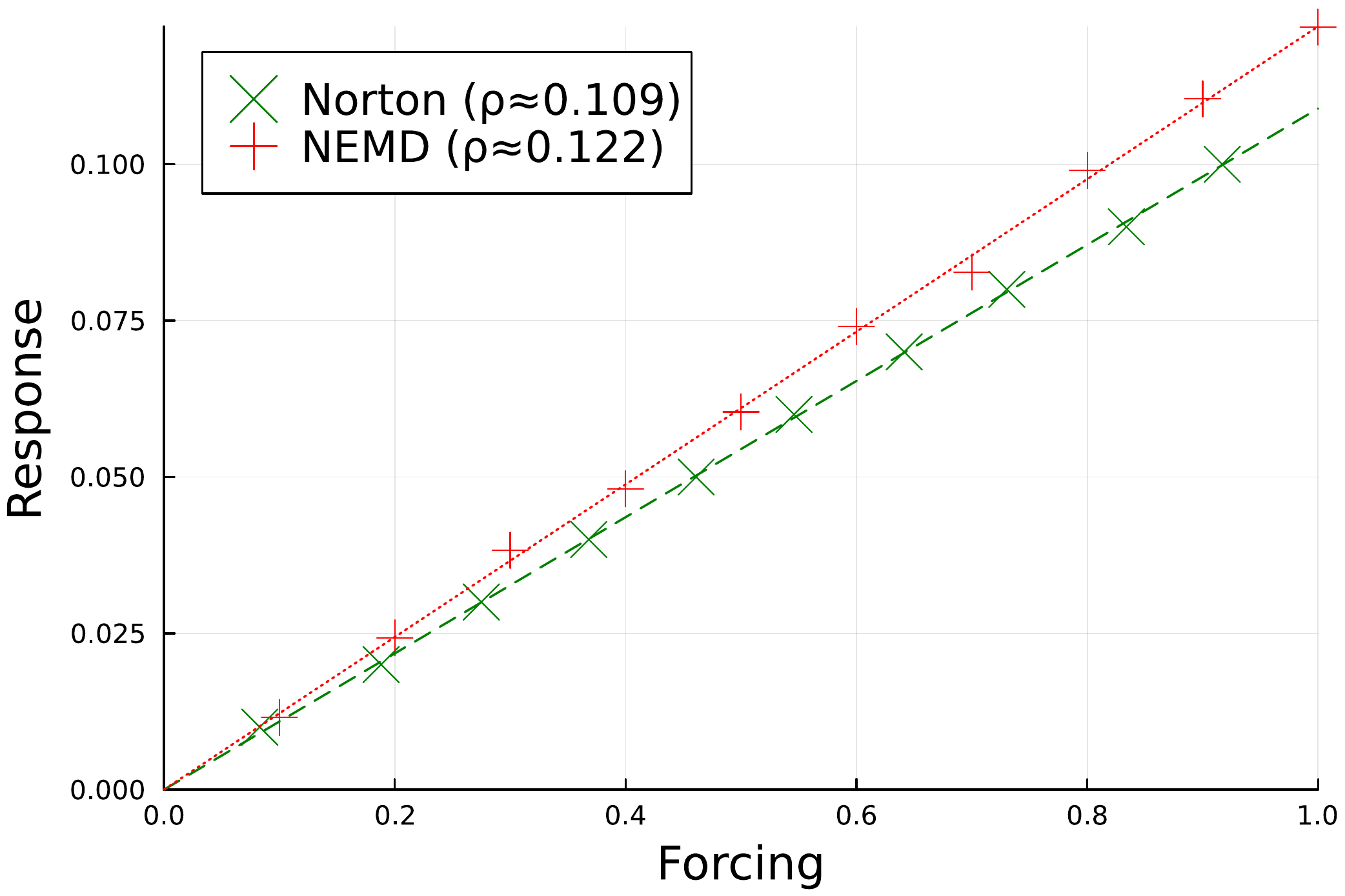}
    \caption{Response profiles in the linear regime for the NEMD and Norton mobility dynamics. The slope of the linear regression lines correspond to the estimated transport coefficient. Left: color drift. Right: single drift.}
    \end{figure}
    \label{fig:mobility_linear}

    In Figure~\ref{fig:mobility_non_linear}, we again look at the color-drift perturbed dynamics, but this time take the system far from equilibrium, so that the response is non-linear. Remarkably, the non-linear responses profiles for the Norton and NEMD systems still coincide for extreme values of the forcing magnitude. This was already formally proven in~\cite{e93} in a deterministic setting.
        \begin{figure}
        \centering
        \includegraphics[width=0.8\textwidth]{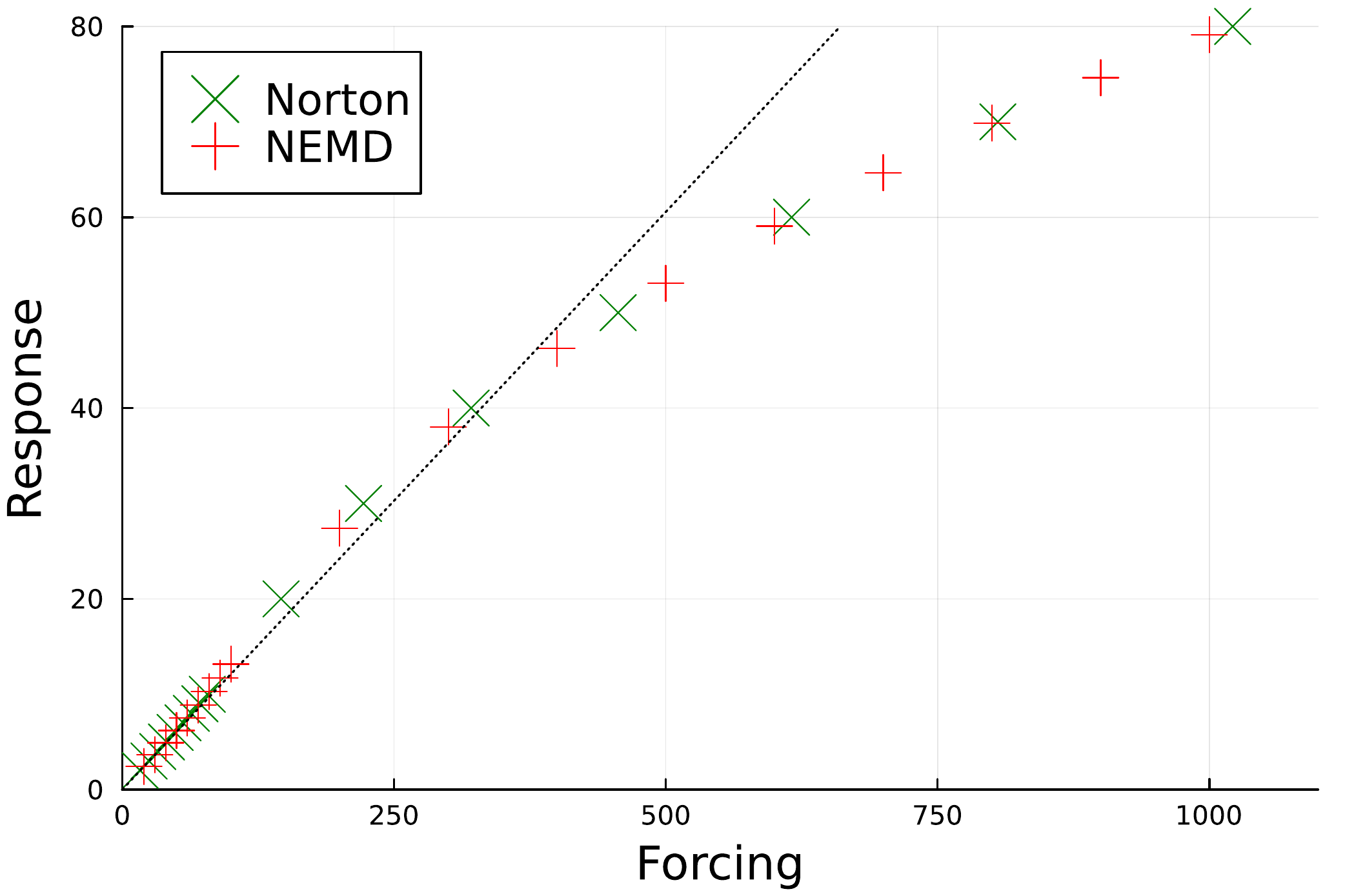}
        \caption{Non-linear response profiles for the NEMD and Norton mobility dynamics with color drift. The responses coincide even in the large perturbation regime.}
        \label{fig:mobility_non_linear}
    \end{figure}

    \paragraph{Shear viscosity computations}
    In Figures~\ref{fig:sv_linear} to~\ref{fig:sv_sinusoidal}, we perform an experiment similar to the one done to estimate the mobility, for a system subject to a longitudinal forcing modulated in intensity by a transverse profile as described in Section~\ref{subsec:langevin_tc}, for the three forcing profiles considered in~\cite{js12}. The systems were simulated under the same conditions as for the mobility, except for the temperature  and particle density which were respectively chosen to be~$T=0.8$ and~$\rho=N/(L_xL_yL_z)=0.7$.

    More precisely, Figure~\ref{fig:sv_linear} 
 presents the response obtained for a piecewise linear forcing profile ~$f(y)=4\1_{0\leq y < L_y/2}(y-L_y/4)/L_y + 4\1_{L_y/2\leq y < L_y}(3L_y/4-y)/L_y)$, Figure~\ref{fig:sv_constant} for a piecewise constant forcing profile~$f(y)=1-2 \1_{0<y\leq L_y/2}$, and Figure~\ref{fig:sv_sinusoidal} for a sinusoidal forcing profile~$f(y)=\sin(2y\pi/L_y)$. In each case, the linear response regime is plotted on the left-hand side, and the non-linear regime is plotted on the right. Linear regression lines whose slopes are given in the legend are superimposed to the data points. Each time, both the linear and non-linear responses for usual NEMD and Norton dynamics perfectly agree.

\paragraph{Consistency in the thermodynamic limit.}
 We now turn to investigating the behavior of estimators of the shear viscosity in the thermodynamic limit~$N\to\infty$.
 In order for the thermodynamic limit of the viscosity to be well-defined, the quantity~$F_1/U_1 -\gamma_x$ must scale asymptotically as~$N^{-2/3}$. Indeed, as the system size increases, the length~$L_y$ scales as~$(N/\rho)^{1/3}$, which, plugged in the expression\[\nu=\rho\left(\frac{F_1}{U_1} - \gamma_x\right)\left(\frac{L_y}{2\pi}\right)^2\]
implies the claimed scaling for~$F_1/U_1 -\gamma_x$. This discussion determines how to fit the asymptotic behavior of~$F_1/U_1$ for increasing system sizes (see Figure~\ref{fig:thermo_limit}).

 In Figure~\ref{fig:thermo_limit}, we examine the behavior of the response for increasing sizes of the system. Estimations of~$U_1$ were performed at roughly equivalent state points~$\eta = 0.3$ and~$r=0.1$ in the linear response regime using estimators of the form~\eqref{eq:discr_estimators_ct}. We find in particular that the scaling law~$F_1/U_1 -\gamma_x \sim N^{-2/3}$ is obeyed, and that the estimates of the prefactor~$C$ coincide for the NEMD and Norton methods. The agreement between the two methods is excellent for all values of~$N$. 

    \begin{figure}
        \centering
        \includegraphics[width=0.49\textwidth]{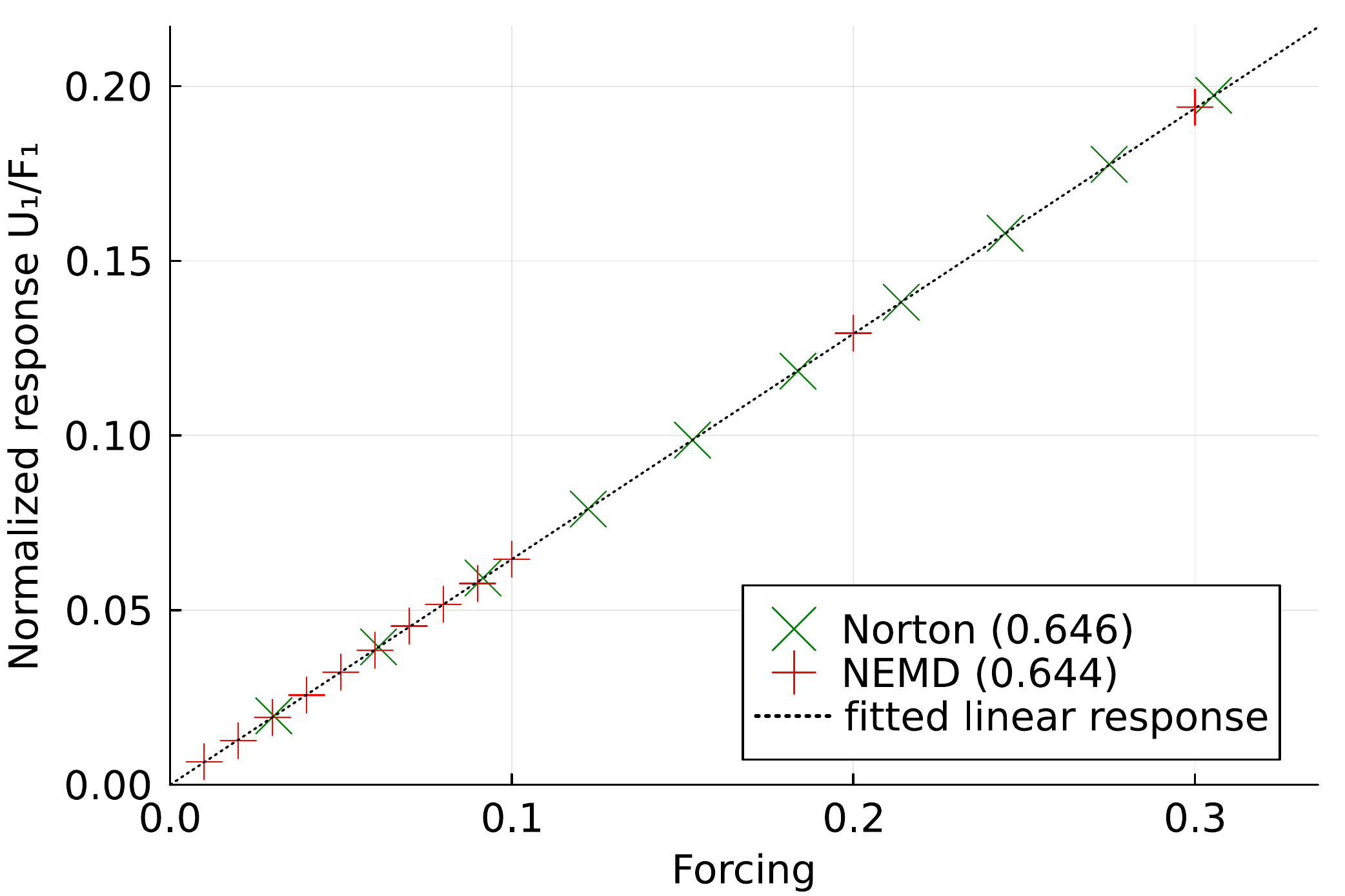}
        \includegraphics[width=0.49\textwidth]{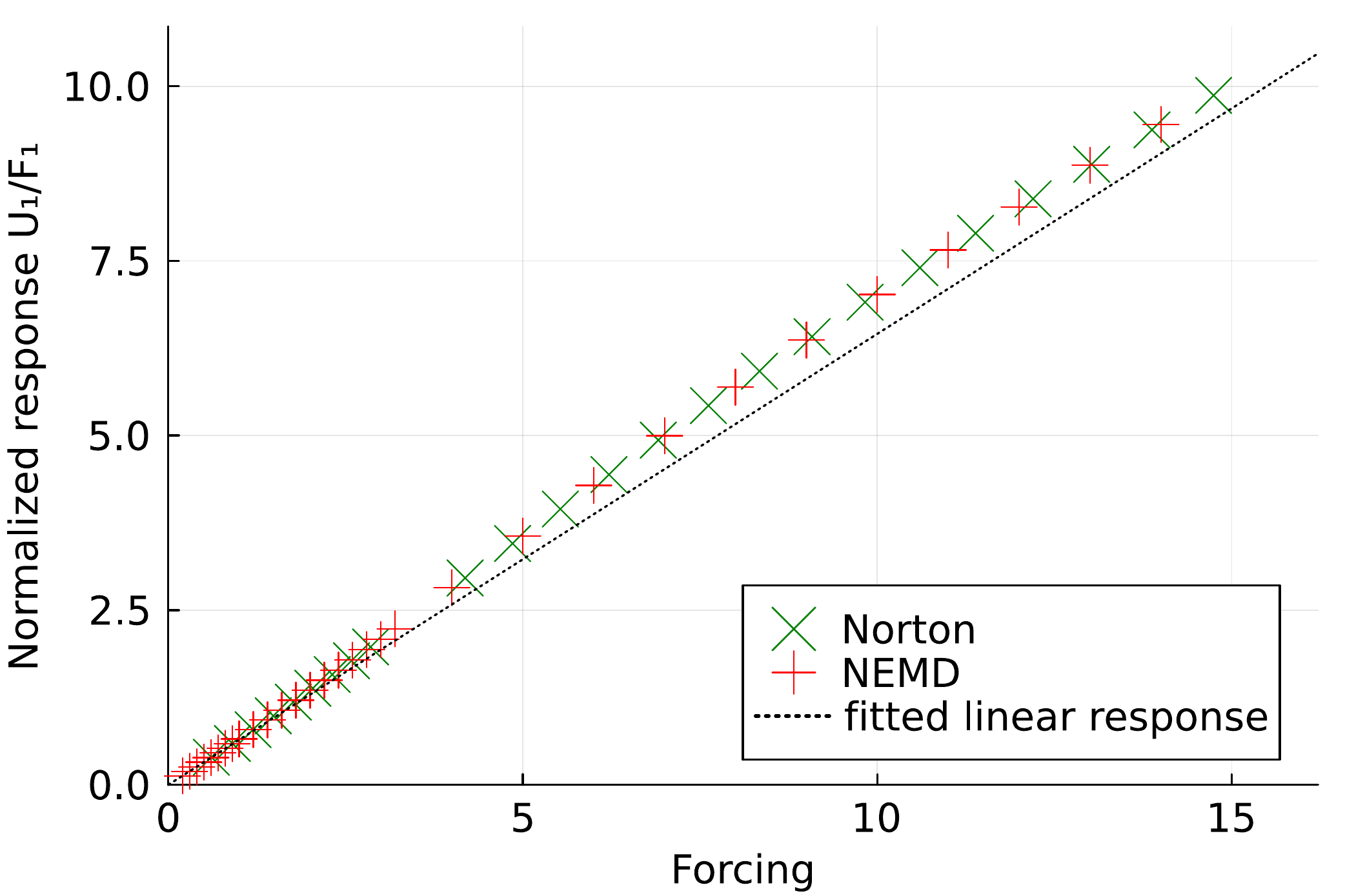}
        \caption{Non-equilibrium normalized Fourier response for shear viscosity computations with a piecewise linear forcing profile.}
        \label{fig:sv_linear}
    \end{figure}

    \begin{figure}
        \centering
        \includegraphics[width=0.49\textwidth]{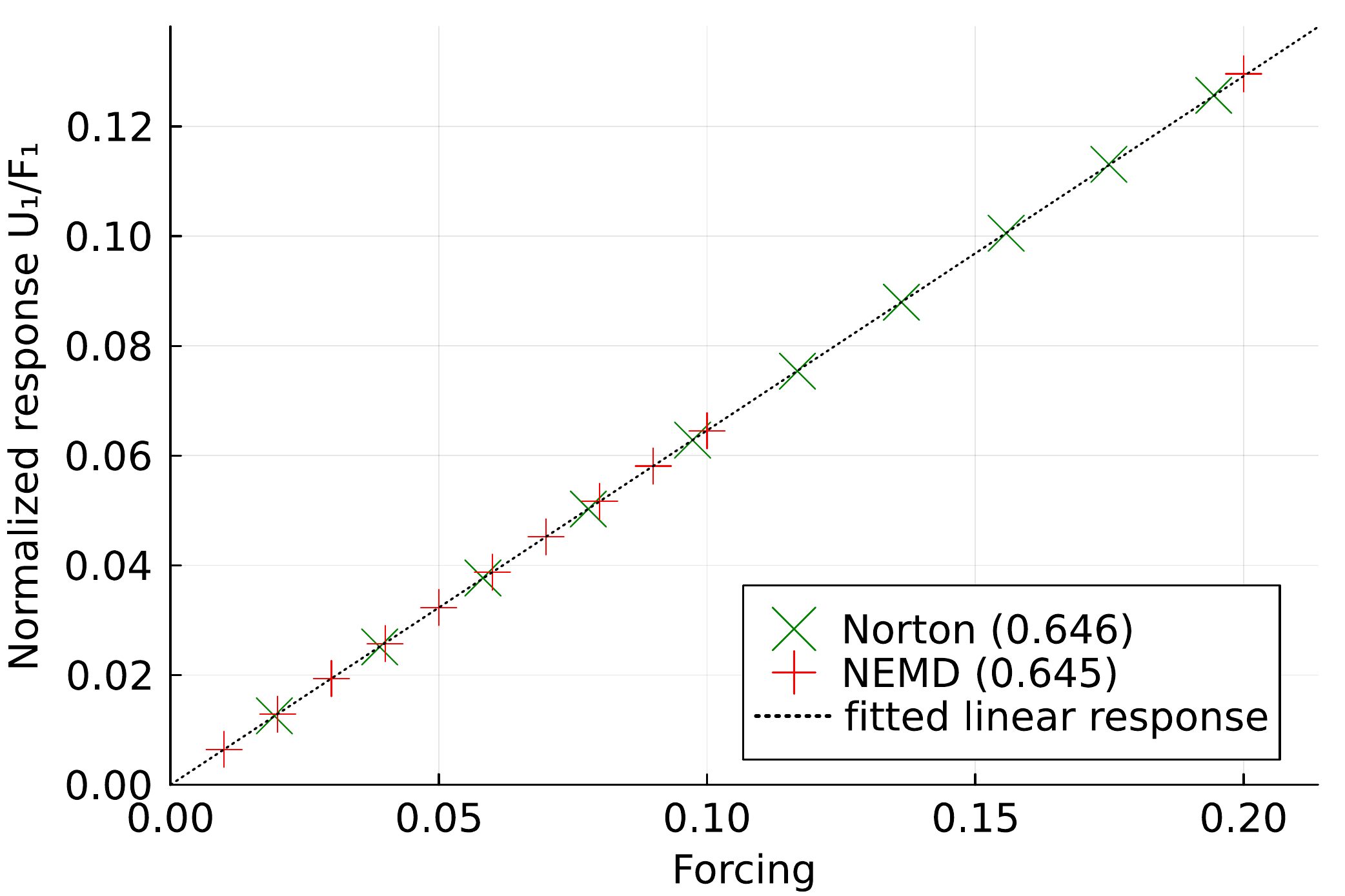}
        \includegraphics[width=0.49\textwidth]{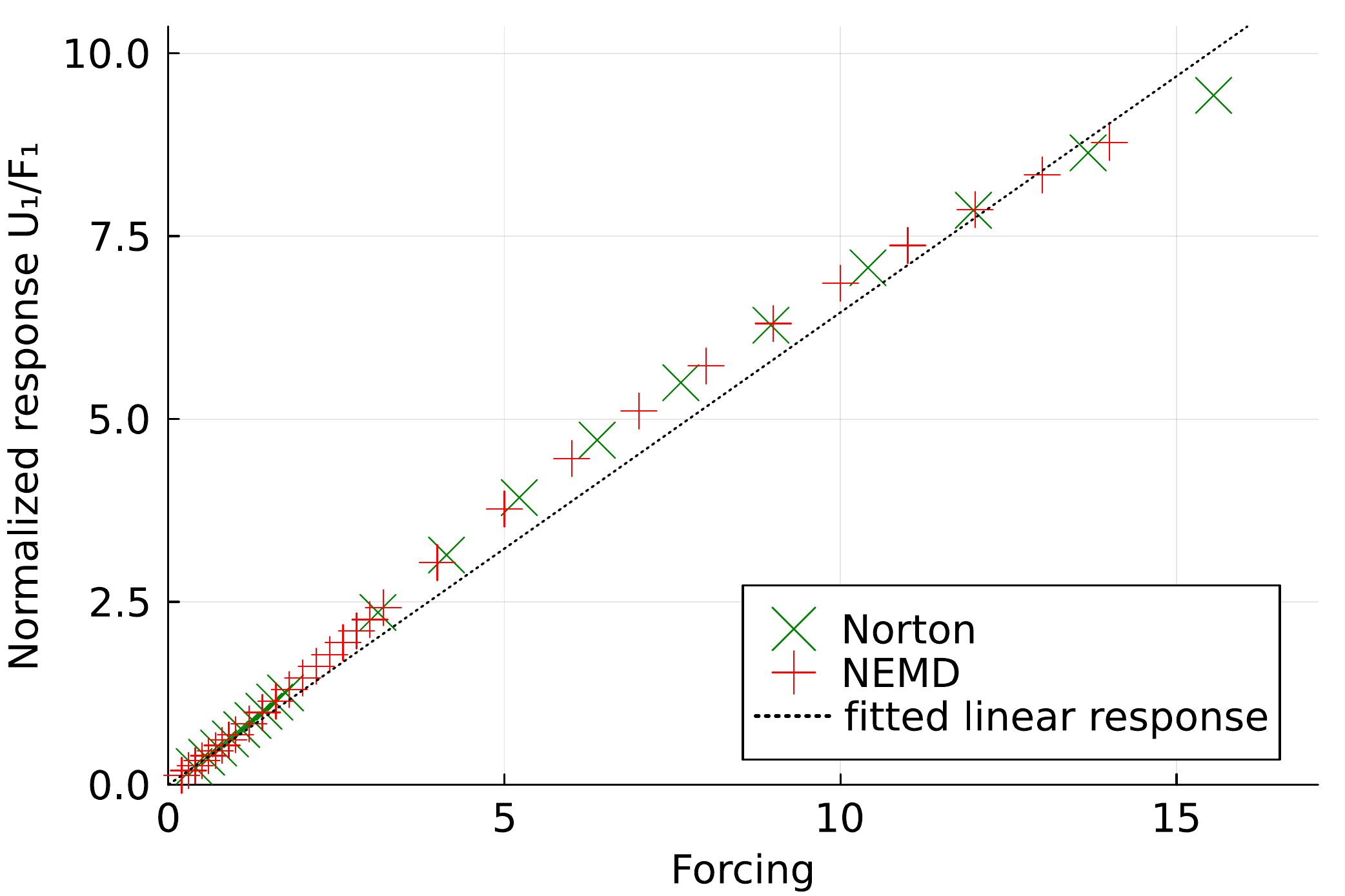}
        \caption{Non-equilibrium normalized Fourier response for shear viscosity computations with a piecewise constant forcing profile.}
        \label{fig:sv_constant}
    \end{figure}
        \begin{figure}
        \centering
        \includegraphics[width=0.49\textwidth]{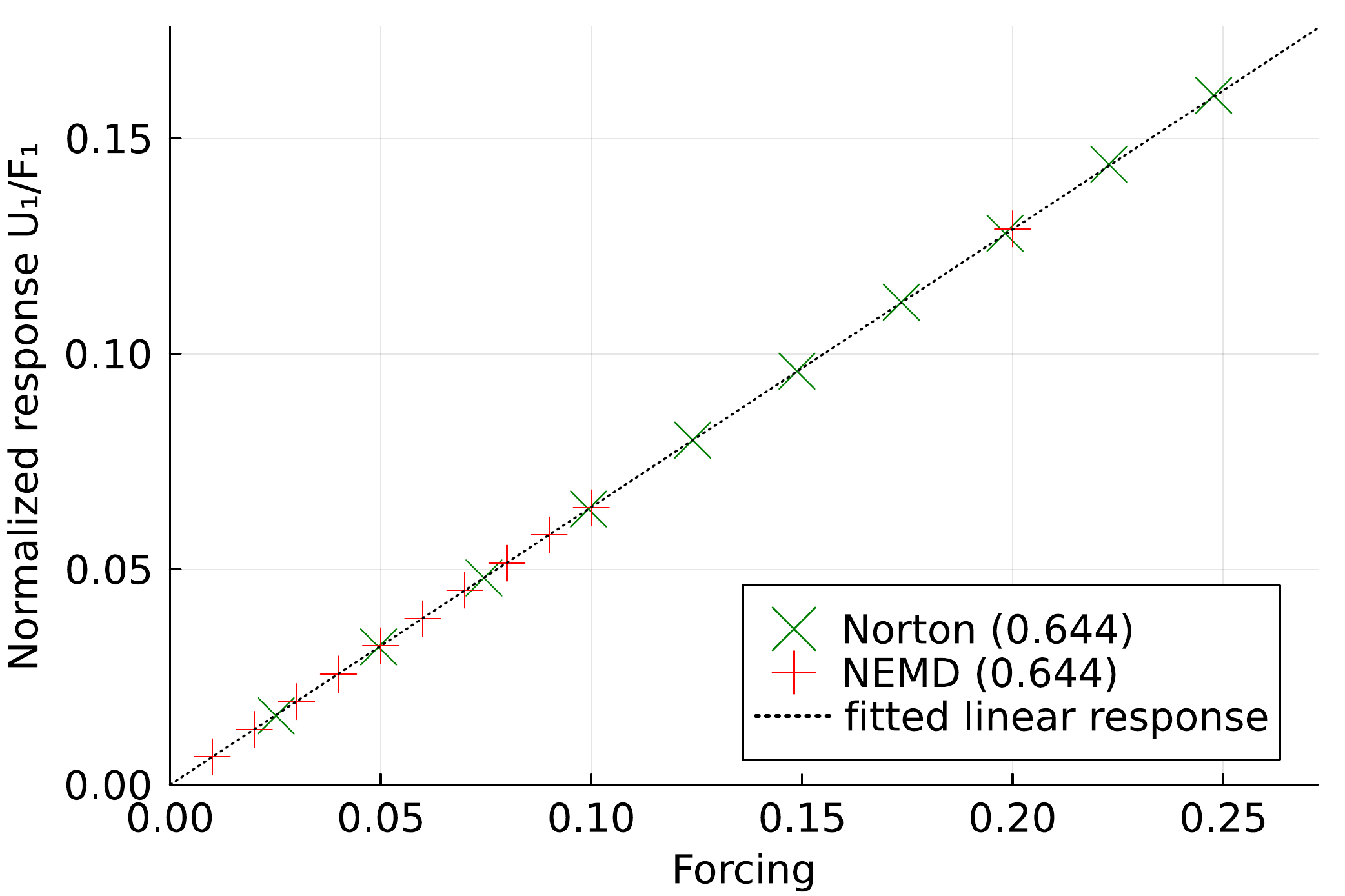}
        \includegraphics[width=0.49\textwidth]{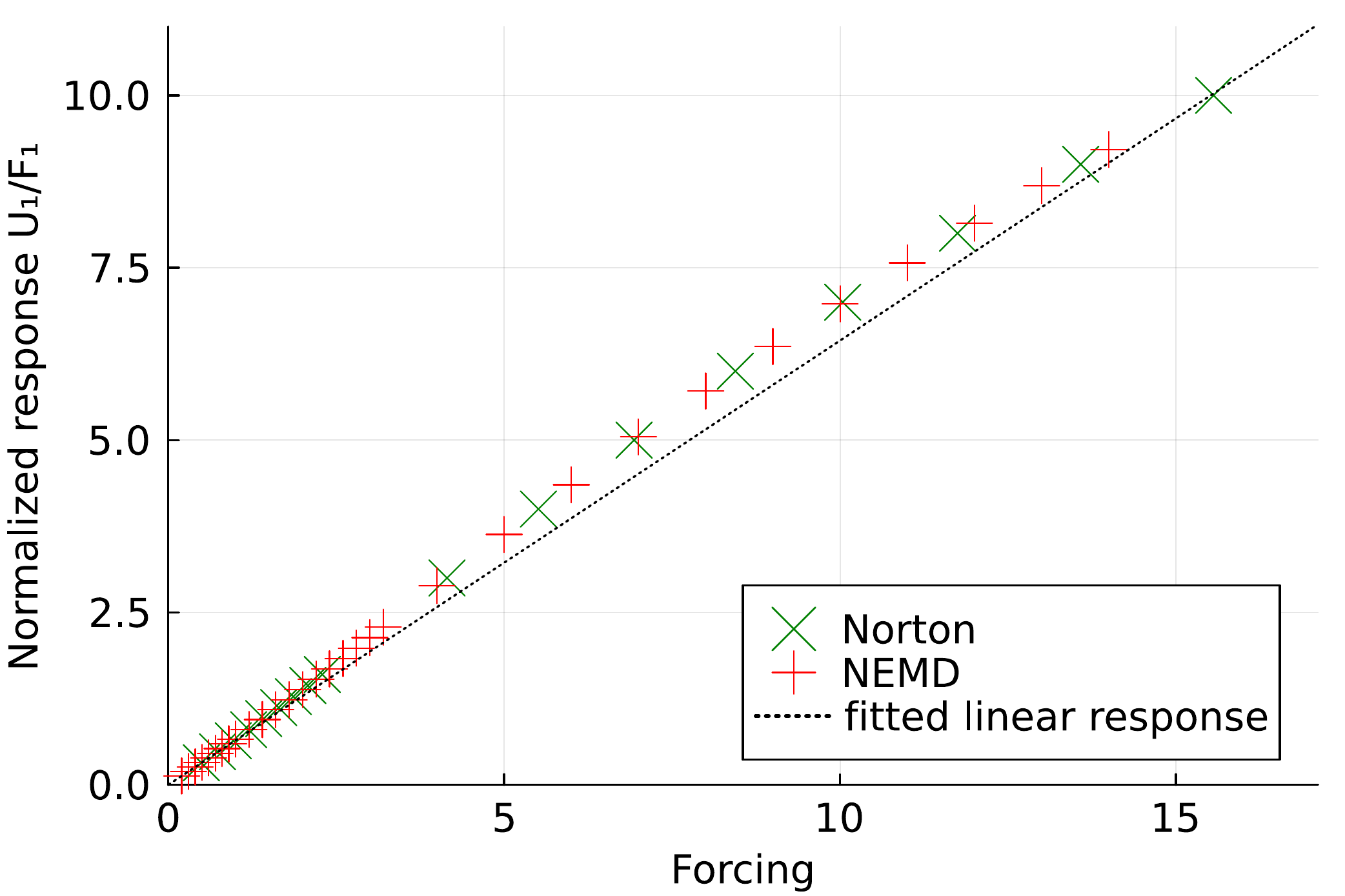}
        \caption{Non-equilibrium normalized Fourier response for shear viscosity computations with a sinusoidal forcing profile.}
        \label{fig:sv_sinusoidal}
    \end{figure}
\end{subsection}
    
    \begin{subsection}{Concentration properties in the thermodynamic limit}\label{subsec:numerical_thermo_limit}

    We next investigate the concentration rates of~$R$ and~$\lambda$ as a function of the system size~$N$, respectively in the canonical equilibrium and Norton ensembles, in the case of a sinusoidal shear forcing profile. In Figure~\ref{fig:histograms}, we show equilibrium distributions of~$R$ and~$\lambda$ for increasing values of~$N$. By \``equilibrium\", it is meant here that the NEMD dynamics is run for~$\eta=0$ and the Norton dynamics for~$r=0$.
    Note first that both distributions are centered around~0 and symmetric. Moreover, they concentrate around~0 as the system size~$N$ increases. The concentration rate however differs for the two dynamics. In the canonical ensemble, the usual concentration rate~$N^{-1}$, consistent with some form of spatial averaging of an intensive quantity, is observed. Indeed, the rate~$N^{-1}$ is the one dictated by a central limit theorem, which makes sense on an intuitive level for a system displaying only short-range interactions, using a decomposition of the full system into weakly correlated subsystems, each of them contributing additively to the total flux.
    
    On the other hand, the concentration rate as a function of the system size in the Norton ensemble appears to be faster. This is confirmed in Figure~\ref{fig:eq_vars}, which shows that the second moment of~$\lambda$ decays as~$N^{-5/3}$. The increase in the rate is presumably due to the coupling introduced by the constant-flux condition, which introduces global correlations in the systems, and hence some form of long range interactions, which are known to lead to significant changes of phenomenology in some situations.

    \begin{figure}
        \centering
        \includegraphics[width=0.49\linewidth]{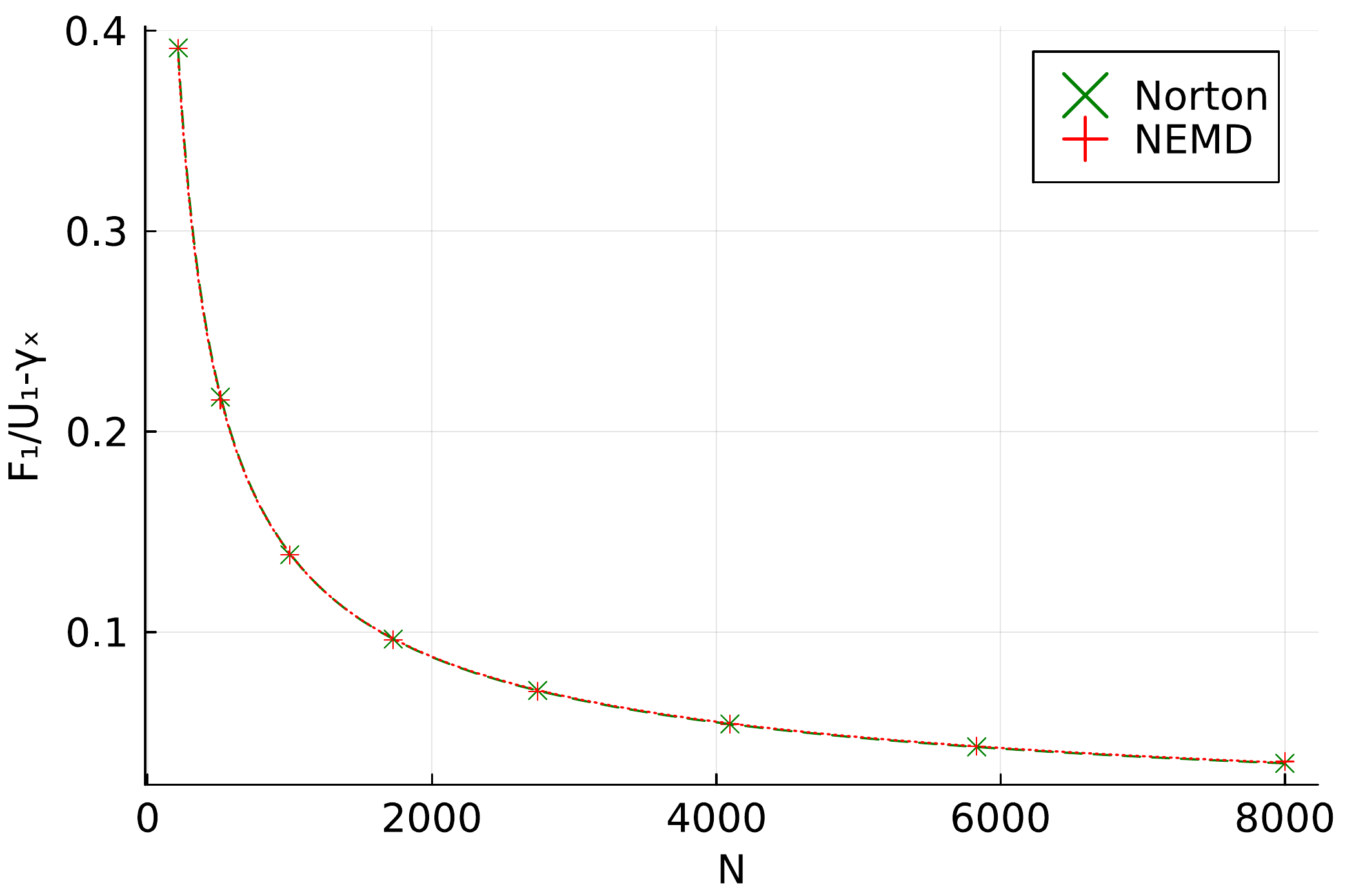}
        \includegraphics[width=0.49\linewidth]{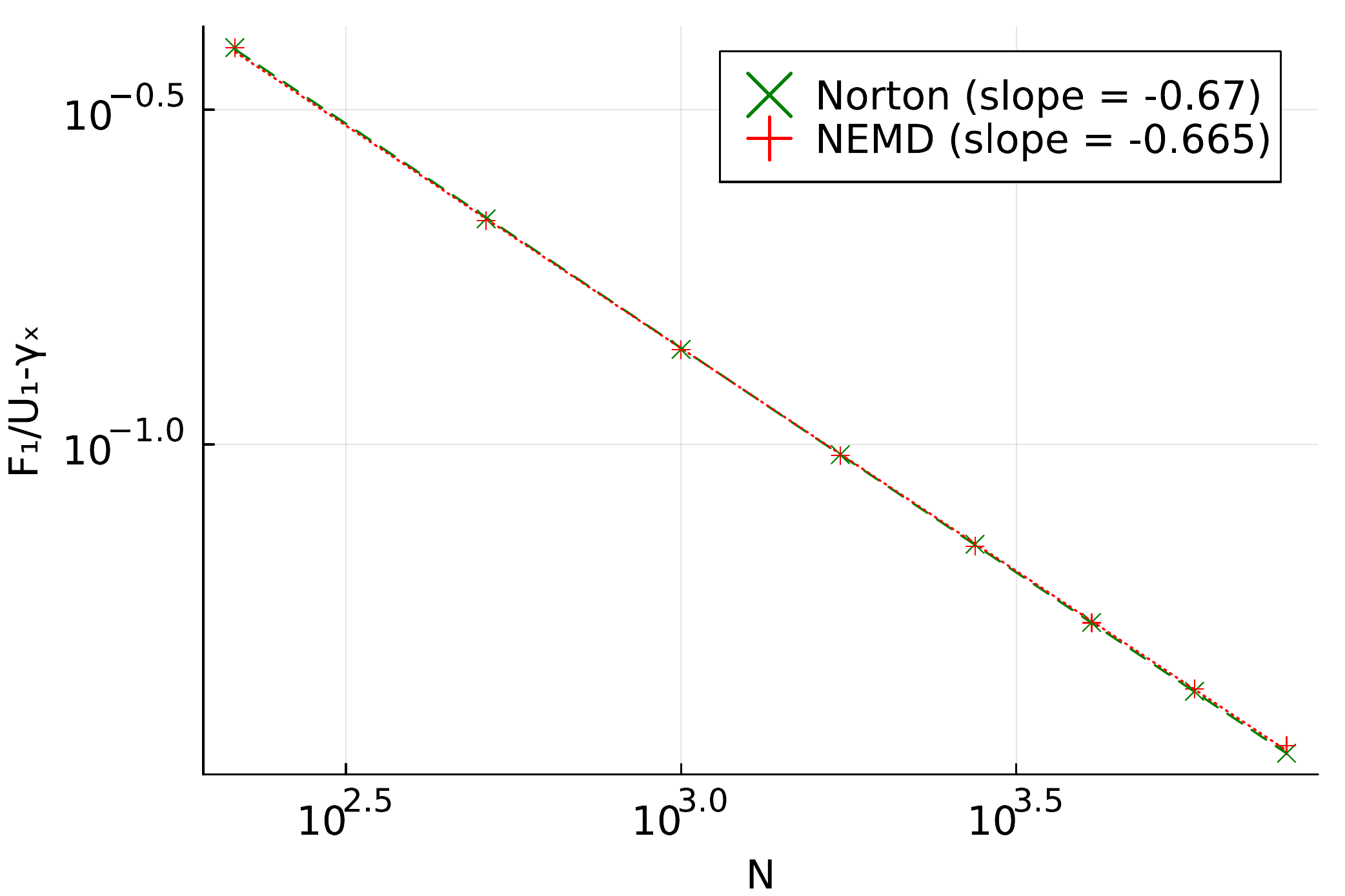}
        \caption{Behavior of estimators for~$U_1/F_1-\gamma_x$ in the large~$N$ limit. The simulation results, indicated by markers, are fitted by~$C N^{-\alpha}$ with~$C\approx 14$ and~$\alpha\approx0.66$ with a linear regression of~$\log(U_1/F_1-\gamma_x)$ as a function of~$\log N$. The right plot corresponds to the left plot in log-log coordinates.}
        \label{fig:thermo_limit}
    \end{figure}

        \begin{figure}
        \centering
        \includegraphics[width=0.49\linewidth]{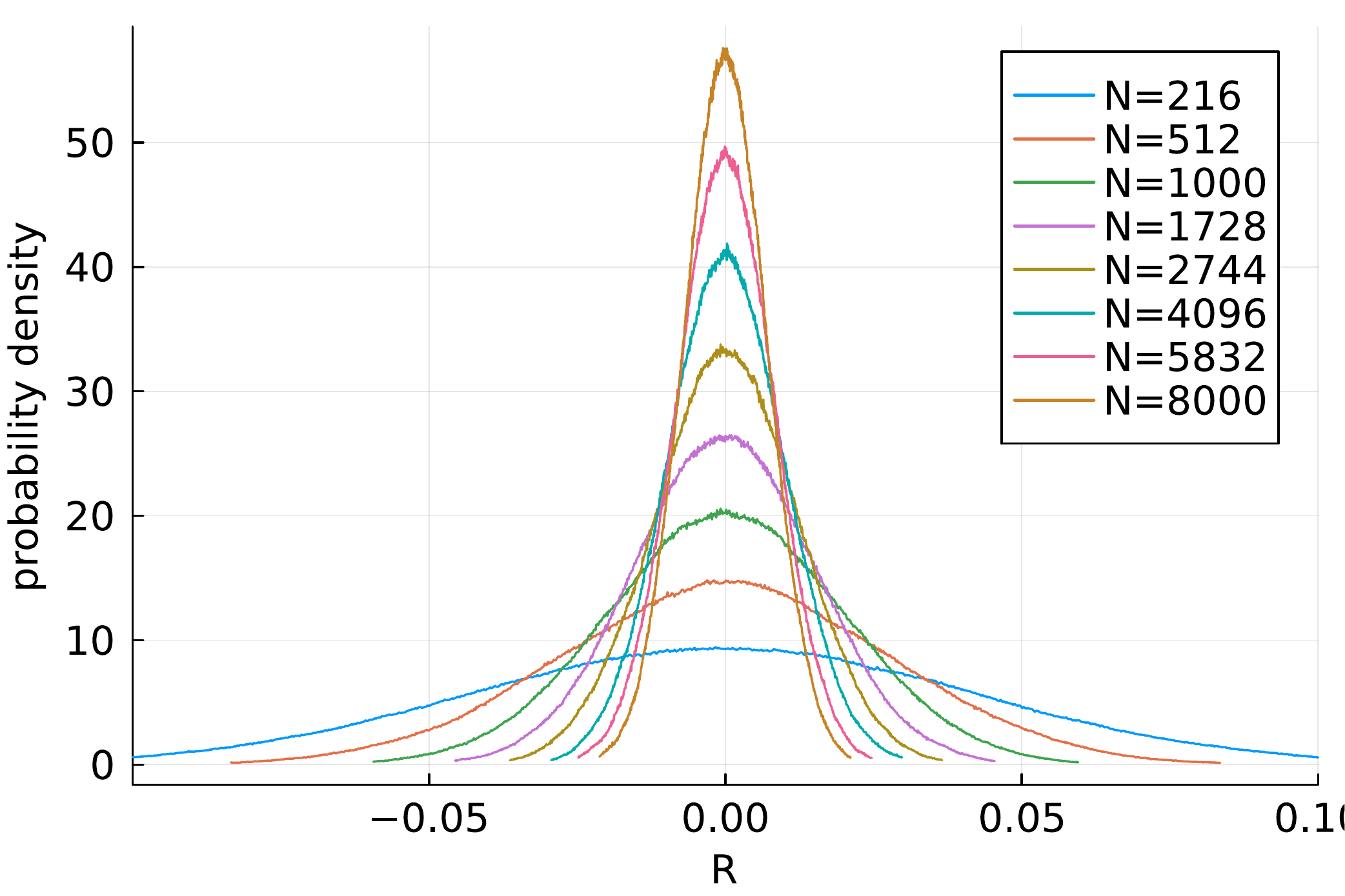}
        \includegraphics[width=0.49\linewidth]{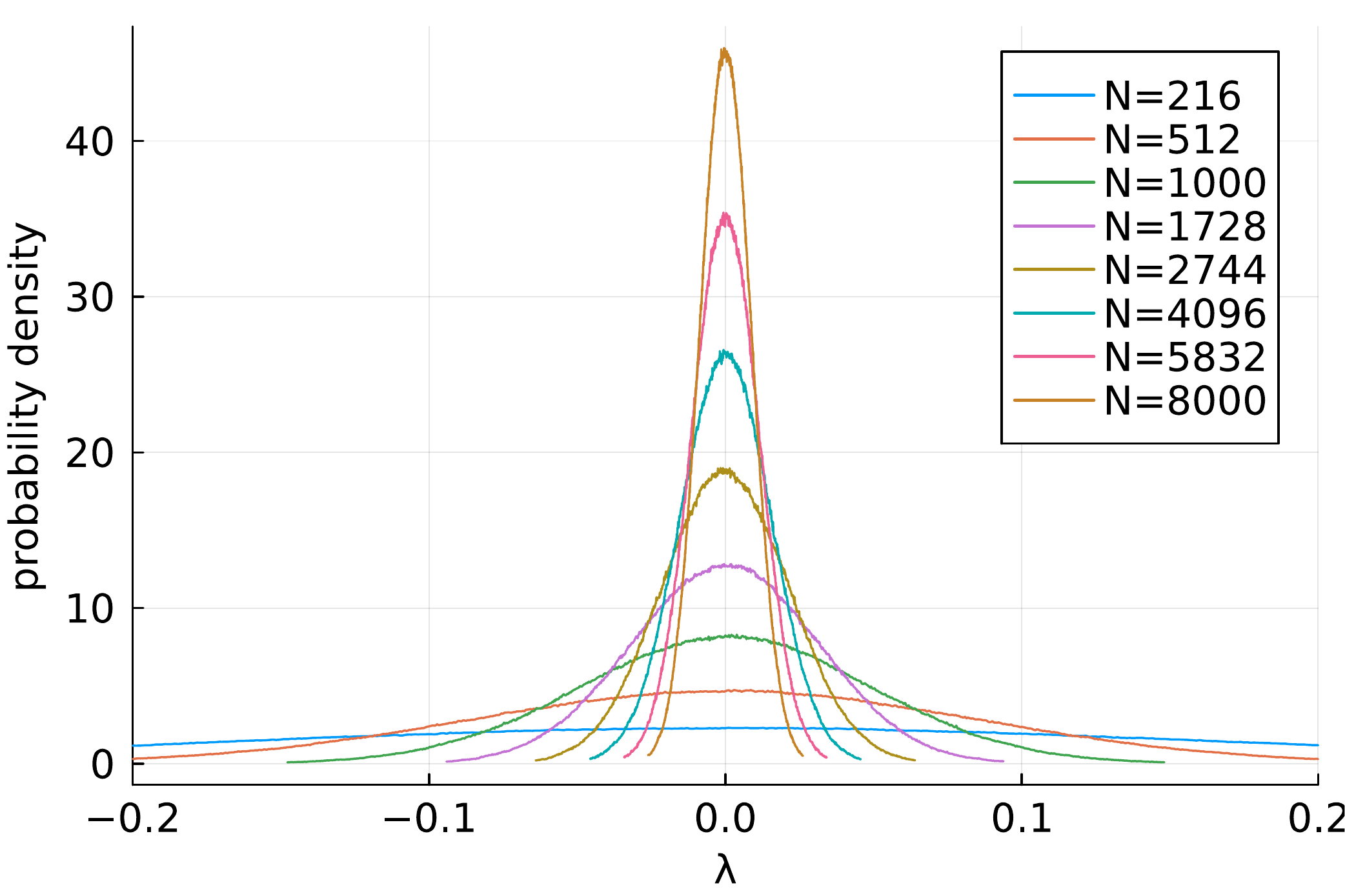}
        \caption{Histograms of forcing and response in equilibrium ensembles. Left: standard equilibrium Langevin dynamics~($\eta=0$). Right: Langevin--Norton dynamics at zero flux~($r=0$).}
        \label{fig:histograms}
    \end{figure}

    \begin{figure}
        \centering
        \includegraphics[width=0.8\linewidth]{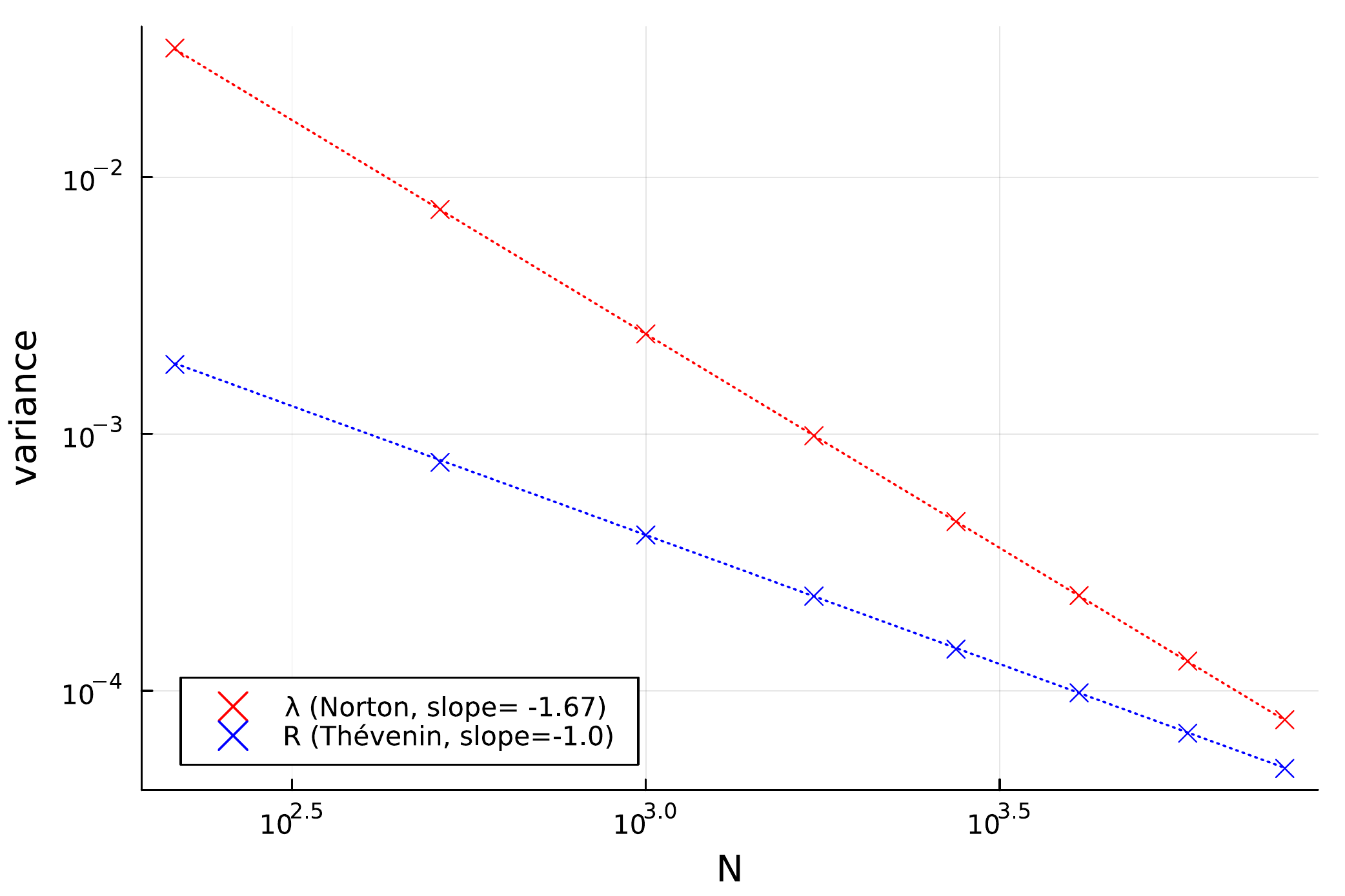}
        \caption{Variance of~$\lambda$ and~$R$ in the Norton and canonical equilibrium ensembles.}
        \label{fig:eq_vars}
    \end{figure}
    \end{subsection}

    \begin{subsection}{Asymptotic variance}\label{subsec:numerical_av}
    We finally assess the numerical efficiency of the Norton method. NEMD estimators of the form~\eqref{eq:discr_estimators_ct} are ergodic averages of~$R$ for trajectories of the discrete-time Markov chain defined by the numerical scheme, and the central limit theorem for Markov chains implies that the variance is asymptotically~$\eta^{-2}N_{\mathrm{iter}}^{-1}\sigma^2_{\Dt}(R)$, where~$\sigma^2_{\Dt,\eta}(R)$ is the asymptotic variance of~$R$ under the scheme's stationary measure~$\pi_{\Delta t,\eta}$:
    \begin{equation}
        \label{eq:discr_av_rhohat}
        \sigma^2_{\Delta t,\eta}(R) = \Var_{\pi_{\Delta t,\eta}}(R) + 2\sum_{k=1}^\infty \Cov_{\pi_{\Delta t,\eta}}\left(R(q^k,p^k)R(q^0,p^0)\right) = \Var_{\pi_{\Delta t,\eta}}(R)\Theta_{\pi_{\Delta t,\eta}}(R),
    \end{equation}
    where 
    \begin{equation}
        \label{eq:discr_correlation_time}
        \Theta_{\pi_{\Delta t,\eta}}(R)=\left[1 + 2\sum_{k=1}^\infty \frac{\Cov_{\pi_{\Delta t,\eta}}\left(R(q^k,p^k)R(q^0,p^0)\right)}{\Var_{\pi_{\Delta t,\eta}}(R)}\right]
    \end{equation}
    is the number of correlation steps~$R$ for stationary initial distribution, and~$\mathrm{Var}_{\pi_{\Delta t,\eta}}(R)$ denotes the centered second moment of~$R$ under~$\pi_{\Delta t,\eta}$. It can be proved for standard NEMD dynamics~\cite{ls16} that
    \[\sigma^2_{\Delta t,\eta}(R) = \sigma^2_\eta(R)/\Delta t + \mathrm{O}(1),\]
    where $\sigma_\eta^2(R)$ is the asymptotic variance~\eqref{eq:nemd_av_asymptotic} of the continuous process.

    For the Norton estimator in~\eqref{eq:discr_estimators_ct}, it can be shown, using the delta method and a computation similar to~\eqref{eq:norton_av}, that the variance of trajectory averages of~$R$ is asymptotically~$N_{\mathrm{iter}}^{-1}r^4\E_{\pi_{\Delta t,r}^*}[\lambda]^{-4}\sigma^{2*}_{\Dt,r}(\lambda)$, where~$\pi_{\Dt,r}^*$ denotes the invariant probability measure of the Markov chain~$(\lambda^n)_{n\geq 1}$ defined in~\eqref{eq:norton_discrete_forcing_estimator}, and~$\sigma^{2*}_{\Dt,r}(\lambda)$  is the associated asymptotic variance for~$\lambda$. Again, we write the asymptotic variance as the product of the centered second moment under the steady-state~$\Var_{\pi_{\Delta t,r}^*}(\lambda_0)$ and the number of correlation steps~$\Theta_{\pi_{\Delta t,r}^*}(\lambda)$.

    As the number of particles~$N$ grows to infinity, and for equivalent state points~$r = \eta\ct$, say in the linear response regime, the numerical results of Section~\ref{subsec:numerical_thermo_limit} show that the contribution of the stationary centered second moment will be asymptotically smaller for Norton estimators than for their NEMD counterparts, owing to the fast concentration rate of~$\lambda$ in the Norton ensemble. This is reason to suspect that Norton estimators of the transport coefficient may have lower asymptotic variance for large enough~$N$, as long as the scaling of correlation times for~$\lambda$ in the Norton ensemble does not cancel this effect. To assess whether this is the case,  we plot in Figure~\ref{fig:shear_acs} autocorrelation functions for equilibrium trajectories of~$\lambda$ in the Norton ensemble and~$R$ in the canonical ensemble, for several values of the system size~$N$, again in the case of a sinusoidal shear forcing profile. These correspond to the summand in~\eqref{eq:discr_correlation_time}, plotted as a function of the physical simulation time. We observe that the correlation time is roughly independent of~$N$. In fact, the autocorrelation profile itself barely changes with increasing system sizes.
    
    Furthermore, we observe that the correlation time is much smaller in the Norton ensemble than in the NEMD ensemble. This suggests that even for moderate system sizes, the asymptotic variance for estimators of the linear response should be smaller for Norton systems than for their NEMD counterparts, owing to smaller correlation times. We verify this intuition in Figure~\ref{fig:sv_avs} for a fixed system of~$N=1000$ particles, in the context of shear viscosity computations. The asymptotic variance for estimators of the Fourier linear response~$U_1$ is plotted as a function of the forcing magnitude, for the various forcing profiles considered in Section~\ref{subsec:numerical_consistency}. Note that the asymptotic variances indeed scale as~$\eta^{-2}$ for NEMD simulations and~$r^{-2}$ for Norton dynamics, at least for small values of~$|r|$ and~$|\eta|$. Interestingly, we find that Norton estimators have lower variance in each situation. We expect that this difference will be more pronounced for larger systems.

    \begin{figure}
        \centering
        \includegraphics[width=0.49\linewidth]{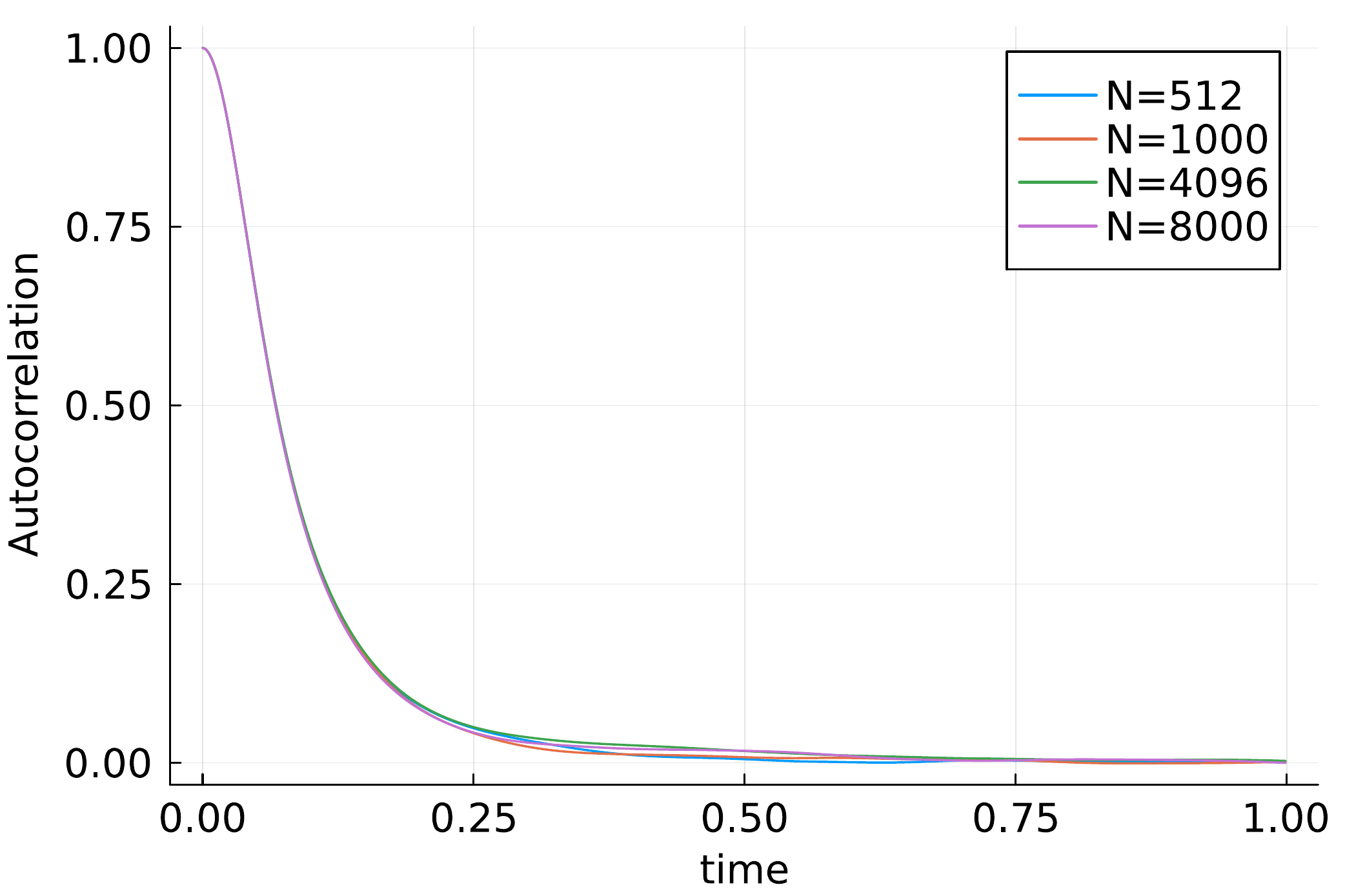}
        \includegraphics[width=0.49\linewidth]{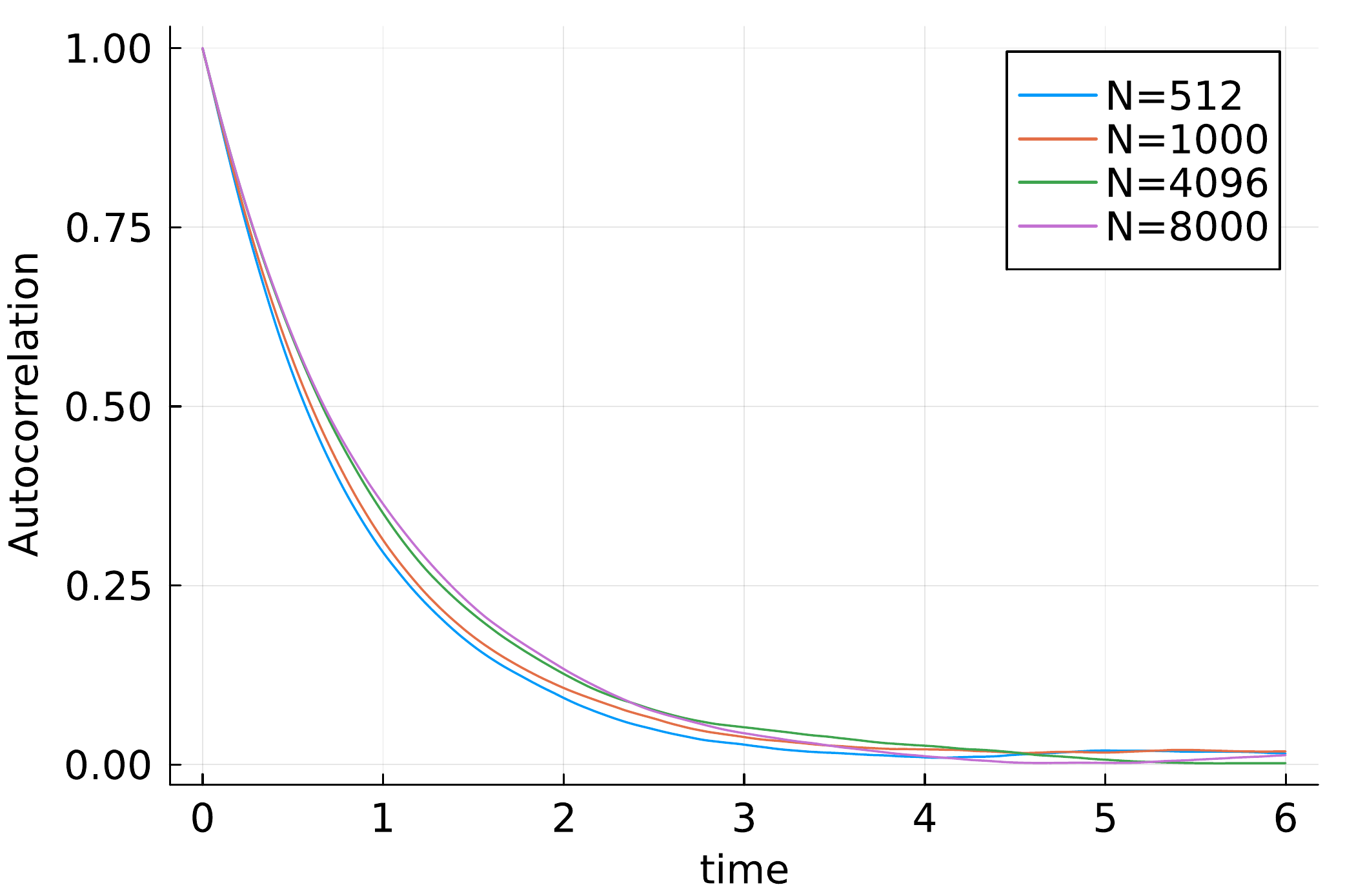}
        \caption{Autocorrelation functions at equilibrium for the shear-viscosity Fourier response $R$ (given by~\eqref{eq:shear_response}). Left: autocorrelation function of~$\lambda$ for the Norton dynamics on $R^{-1}\{0\}$. Right: autocorrelation function of $R$ in standard equilibrium simulations.}
        \label{fig:shear_acs}
    \end{figure}

    \begin{figure}
        \centering
        \includegraphics[width=0.8\linewidth]{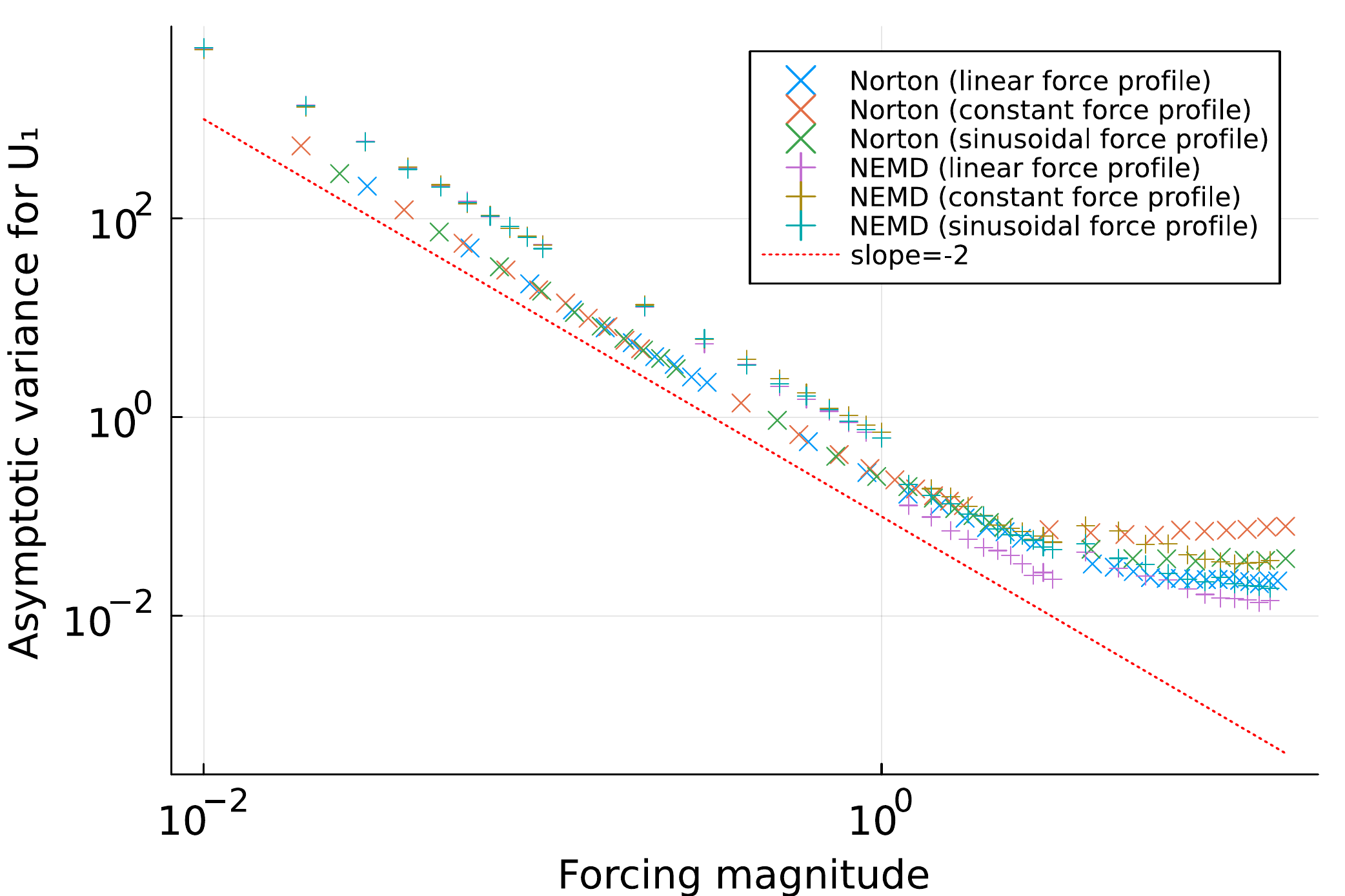}
        \caption{Asymptotic variance for estimators of the Fourier response~$U_1$ as a function of the perturbation magnitude, in log-log coordinates. The expected scaling line for small values of the forcing or response is plotted in red. }
        \label{fig:sv_avs}
    \end{figure}
    
    \end{subsection}
    
\section{Perspectives for future work}\label{sec:perspective}
We list in this final section some of the many open questions raised by the stochastic Norton approach, both on the theoretical and numerical sides. Certain points are currently being tackled, but a lot remains to be done, and we hope that this work will encourage other researchers to study Norton dynamics. We classify issues to address in three families.

\paragraph{Analysis of the continuous dynamics.}
A first task is to show that Norton dynamics are well-posed and well-behaved. As such results will undoubtedly constitute necessary tools to investigate the properties of Norton ensembles from a rigorous standpoint, they are of the utmost importance. We identify the following questions:
\begin{itemize}
    \item Obtain sufficient conditions on the reference dynamics,~$F$ and~$R$, for the Norton dynamics to be well-posed on arbitrarily large time horizons. 
    \item Prove the existence and uniqueness of a Norton steady-state.
    \item Show pathwise ergodicity of the Norton dynamics.
    \item Obtain bounds on the rate of convergence of arbitrary initial conditions to the Norton steady-state.
\end{itemize}
In view of conditions such as~\eqref{eq:norton_controllability_condition}, it is clear that not all pairs~$(F,R)$ give rise to well-posed Norton dynamics. From a macroscopic point of view, the averages of~$F$ and~$R$ correspond to conjugate variables, hence we expect any reasonable microscopic realization to appropriately relate the associated observables.
\paragraph{Theoretical questions.}
We list here various physically motivated questions, which would in particular establish that the Norton method provides a viable way to compute transport coefficients.
\begin{itemize}
    \item Derive equivalence of ensemble results between the Norton and canonical ensemble at equilibrium (i.e. $r=0$ and~$\eta = 0$, respectively), in the thermodynamic limit.
    \item Prove the existence of the linear response~\eqref{eq:def_norton_transport_coeff} for all systems of finite size, and in fact develop a theory of linear response theory for Norton ensembles. This would include relating the transport coefficient to fluctuations in the Norton equilibrium ensemble, analogously to standard Green--Kubo formul\ae\ in the canonical ensemble. This approach was formally carried out in~\cite{em85}, for the special case of electrical conductivity or mobility, in a deterministic setting. Carrying out this development would in particular require understanding, for small~$|r|$, the relation of the Norton equilibrium steady-state to the invariant measure of the reference process~\eqref{eq:thevenin_dynamics}, as well as to the nature of the forcing~$F$ and the response~$R$.
    \item Show the consistency of the Norton approach, by proving that the transport coefficients obtained with NEMD and Norton dynamics have the same thermodynamic limit, for bulk forcings. This can be considered as some form of equivalence of ensembles in a perturbative nonequilibrium setting.
    \item Derive a result on the equivalence of ensembles for the Norton and NEMD approaches beyond the linear response regime, for equivalent state points (i.e. the average forcing experienced in Norton dynamics coincides with the fixed forcing in NEMD simulations).
\end{itemize}
\paragraph{Numerical analysis of Norton dynamics.}
A last set of issues to address is related to properties of Norton dynamics which are of practical interest for the efficient and accurate computation of transport coefficient in realistic systems. A crucial point is to derive clear conditions under which one should prefer the Norton approach to the usual NEMD method. Another point concerns the study of the mathematical properties of the discretized dynamics.
\begin{itemize}
    \item Carry out a rigorous analysis of the concentration rate of~$\lambda$ in the thermodynamic limit.
    \item Explain the shorter correlation times for Norton dynamics compared to the corresponding NEMD dynamics. This and the previous question would ideally yield rigorous asymptotic formulas for both the variance and correlation time in terms of~$\eta$,~$r$ and~$N$.
    \item Perform the numerical analysis of splitting schemes for Norton dynamics. This analysis is a necessary step to derive error bounds relating quantitative properties of the continuous dynamics with their discrete analogs. Useful results would include error estimates at the level of discrete trajectories, or at the level of the invariant measures of the associated Markov chains.
\end{itemize}

\appendix

\section{Proofs of some technical results}
We gather in this appendix the proofs of various technical statements, namely Equation~\eqref{eq:norton_multiple_constraints} in Section~\ref{appendix:multiple_constraints}, Equation~\eqref{eq:norton_time_varying_constraint} in Section~\ref{appendix:time_dependent_constraints}, and Equation~\eqref{eq:color_drift_relation} in Section~\ref{appendix:color_drift}. 
\subsection{Derivation of Norton dynamics in the case of multiple constraints}
\label{appendix:multiple_constraints}
We write here the proof of~\eqref{eq:norton_multiple_constraints}.
We proceed as in Section~\ref{subsec:norton_lambda}, assuming that the forcing process $\Lambda^{\mathbf r}$ may be written as an Itô process before checking a posteriori that this assumption is justified. More precisely, we write
\[\d \Lambda_t^{\mathbf r} = \d \widetilde{\Lambda}_t^{\mathbf r} + \lambda^{\mathbf r}_t \, \d t,\]
where $\widetilde{\Lambda}^{\mathbf r}$ is the martingale part of the Itô decomposition of $\Lambda^{\mathbf r}$.
Applying Itô's formula to the constant-response constraint $R(Y^{\mathbf r}_t) = R(Y^{\mathbf r}_0)=\mathbf{r}$ yields
\begin{equation}\label{eq:ito_formula_multiple_constraints}\nabla R(Y^{\mathbf r}_t)^\intercal\d Y^{\mathbf r}_t + \frac12\nabla^2 R(Y^{\mathbf r}_t) : \d \langle Y^{\mathbf r}\rangle_t = 0.\end{equation}
Identifying martingale increments, the uniqueness of the Itô decomposition implies
\[\d \widetilde{\Lambda}_t^{\mathbf r} = - \left[\nabla R(Y_t^{\mathbf{r}})^\intercal F(Y_t^{\mathbf{r}})\right]^{-1}\nabla R(Y_t^{\mathbf{r}})^\intercal \sigma(Y_t^{\mathbf r}) \,\d W_t,\]
which in turns allows to compute the quadratic covariation term
\[\d \langle Y^{\mathbf r}\rangle_t = \left[\overline{P}_{F,\nabla R} \sigma\sigma ^\intercal \overline{P}_{F,\nabla R}^\intercal\right](Y_t^{\mathbf r})\,\d  t = \Pi_{F,\nabla R,\sigma}(Y_t^{\mathbf r})\, \d t.\]
Using this expression in~\eqref{eq:ito_formula_multiple_constraints} and the uniqueness of the Itô decomposition once again to identify bounded-variation parts, we recover
\[\lambda_t^{\mathbf r} = - (\nabla R^\intercal F)^{-1}\left[\nabla R^\intercal b +\frac12\left(\nabla^2 R^\intercal : \Pi_{F,\nabla R,\sigma}\right)\right] (Y_t^{\mathbf{r}}),\]
which yields~\eqref{eq:norton_multiple_constraints} upon substituting the expression for $\d\Lambda_t^{\mathbf r}$.

\subsection{Derivation of the Norton dynamics in the case of time-dependent constraints}
\label{appendix:time_dependent_constraints}
To prove~\eqref{eq:norton_time_varying_constraint}, we once again proceed as in Section~\ref{subsec:norton_lambda}, writing the Itô decomposition for the forcing process as
\[\d \Lambda_t^{\mathcal R} = \d \widetilde{\Lambda}_t^{\mathcal R} + \lambda^{\mathcal R}_t\,\d t.\]
Applying Itô's formula to the time-dependent constraint~\eqref{eq:norton_time_varying_constraint}, we get
\[\nabla R(Y^{\mathcal R}_t)^\intercal\d Y^{\mathcal R}_t + \frac12\nabla^2 R(Y^{\mathcal R}_t) : \d \langle Y^{\mathcal R}\rangle_t =  \overline{r}_t\,\d t+\widetilde{r}_t\,\d B_t,\]
which, by uniqueness of the Itô decomposition, leads to
\[\d \widetilde{\Lambda}_t^{\mathcal R} = \frac{ \widetilde{r}_t\,\d B_t -\nabla R(Y_t^{\mathcal{R}})\cdot \sigma(Y_t^{\mathcal R}) \,\d W_t}{\nabla R(Y_t^{\mathcal{R}})\cdot F(Y_t^{\mathcal{R}})},\]
allowing to compute the covariation bracket as
\[\d \left\langle Y^{\mathcal R}\right\rangle_t = \Pi_{F,\nabla R,\sigma}(Y^{\mathcal R}_t)\,\d t +\frac{F(Y^{\mathcal R}_t)\otimes F(Y^{\mathcal R}_t)}{\left(\nabla R(Y^{\mathcal R}_t)\cdot F(Y^{\mathcal R}_t)\right)^2}\widetilde{r}_t^2\, \d t.\]
One can then proceed to identify bounded-variation increments, and isolate~$\lambda^{\mathcal R}_t$:
\[\lambda^{\mathcal R}_t = \frac{\displaystyle{\overline{r}_t-\frac12\nabla^2 R(Y^{\mathcal R}_t) : \left[ \Pi_{F,\nabla R,\sigma}(Y^{\mathcal R}_t) +\frac{F(Y^{\mathcal R}_t)\otimes F(Y^{\mathcal R}_t)}{\left(\nabla R(Y^{\mathcal R}_t)\cdot F(Y^{\mathcal R}_t)\right)^2}\widetilde{r}_t^2\right] - \nabla R(Y^{\mathcal R}_t)\cdot b(Y^{\mathcal R}_t)}}{\nabla R(Y^{\mathcal R}_t) \cdot F(Y^{\mathcal R}_t)}.\]
Substituting in the expression for the forcing process finally yields~\eqref{eq:norton_time_varying_constraint}.

\subsection{Derivation of the relation between color/single drift linear responses}
\label{appendix:color_drift}

The proof of~\eqref{eq:color_drift_relation} is taken from unpublished notes by Julien Roussel, see the PhD thesis~\cite{r18}.
We assume for the ease of presentation that~$M=m\mathrm{Id}$ and~$d=3$, and denote by~$\mu$ the equilibrium measure~\eqref{eq:boltzmann_gibbs_measure}. The dynamics under consideration is the standard Langevin dynamics, namely~\eqref{eq:langevin_equation} with~$\eta=0$. Let us define, for $1\leq i,j\leq N$,
\[c_{ij}=\frac{\beta}{m^2} \int_{0}^\infty \E_\mu[p_{i,x,t}p_{j,x,0}]\,\d t.\]
In view of~\eqref{eq:pairwise_potential_type_condition}, upon summing over~$i$ the longitudinal~$p$-components of the SDE~\eqref{eq:langevin_equation} and integrating in time,
\begin{align*}\sum_{i=1}^N p_{i,x,t}&= \sum_{i=1}^N \left[p_{i,x,0}+\int_0^t \left(-\frac{\partial}{\partial q_{i,x}}V(q_s)-\frac{\gamma}{m} p_{i,x,s}\,\d s +\sqrt{\frac{2\gamma}\beta}\,\d W_{i,x,s}\right)\right]\\
    &=\sum_{i=1}^N \left[p_{i,x,0}-\frac{\gamma}{m}\int_0^t  p_{i,x,s}\,\d s +\sqrt{\frac{2\gamma}\beta} W_{i,x,t}\right].\end{align*}
Multiplying by $p_{1,x,0}$ and taking the expectation over trajectories started from canonical initial conditions (so that the Brownian terms vanishes), it follows that
\begin{equation}
    \E_\mu\left[\left(\sum_{i=1}^N p_{i,x,t}\right)p_{1,x,0}\right]=\sum_{i=1}^N\left( \E_\mu\left[p_{i,x,0}p_{1,x,0}\right]-\frac{\gamma}{m}\int_0^t \E_\mu\left[p_{i,x,s}p_{1,x,0}\right]\,\d s\right).
\end{equation}
By the decay properties of the evolution semigroup (obtained for instance by hypocoercive approaches, see e.g.~\cite{v06}, as well as the introduction of~\cite{BFLS22} for a review), the left-hand side converges to~$0$ as~$t\to \infty$, while the integral is well-defined. Since~$p_0$ has diagonal covariance with respect to $\mu$, we get
\begin{equation}
    \sum_{i=1}^N \frac{\gamma}m\int_0^\infty \E_\mu\left[p_{i,x,s}p_{1,x,0}\right]\,\d s=\E_\mu[p_{1,x,0}^2]=\frac{m}{\beta}.
\end{equation}
Equivalently, 
\begin{equation}
\label{eq:color_drift_relation_proof}
\sum_{i=1}^N c_{i1}=\frac{1}{\gamma}.\end{equation}
Using the indistinguishability of the particles,
 \begin{equation}
 \label{eq:cii_cij_relation}
     \forall \,1\leq i,j\leq N,\qquad c_{ii}=c_{11},\qquad c_{ij}=c_{12}.
 \end{equation}
 We can therefore rewrite~\eqref{eq:color_drift_relation_proof} as 
 \begin{equation}\label{eq:c12_c11_relation}c_{12}=\frac{1}{N-1}\left(\frac{1}{\gamma}-c_{11}\right).\end{equation}
This equality can be used to relate the linear responses of the single drift and the color drift.
By the Green--Kubo formula~\cite{rodenhausen}, the transport coefficient for the single drift is given by
\[\ct_{F_{\mathrm{S}}}=c_{11}.\]
For the color drift, we expand, using the Green--Kubo formula,
\[\ct_{F_{\mathrm{C}}}=\frac{\beta}{m^2} \int_0^\infty \E_\mu[\left(F_{\mathrm{C}}\cdot p_t\right)\left(F_{\mathrm{C}}\cdot p_0\right)]\,\d t=\frac1N\left(\sum_{i=1}^N c_{ii} + 2\sum_{1\leq i< j\leq N}(-1)^{i+j}c_{ij}\right)\]
In view of~\eqref{eq:cii_cij_relation}, and using
\[\sum_{1\leq i< j\leq N}(-1)^{i+j}=-\left\lfloor \frac{N}{2}\right\rfloor,\]
which is easily seen by induction, we get
\begin{equation}
    \ct_{F_{\mathrm{C}}}=c_{11}-\frac{2\lfloor N/2 \rfloor}{N(N-1)}\left(\frac1{\gamma}-c_{11}\right),
\end{equation}
which is the claimed identity.
\paragraph{Acknowledgments.}
The authors thank Gary Morris, Julien Roussel and Giovanni Ciccotti for stimulating discussions, and the
members of the project ANR SINEQ (in particular Benjamin Jourdain, Tony
Lelièvre, Stefano Olla and Mathias Rousset) for valuable feedback on preliminary
results presented at the annual meeting of the project. This work
received funding from the European Research Council (ERC) under the
European Union's Horizon 2020 research and innovation programme (project
EMC2, grant agreement No 810367), and from the Agence Nationale de la
Recherche, under grants ANR-19-CE40-0010-01 (QuAMProcs) and
ANR-21-CE40-0006 (SINEQ).

\bibliographystyle{plain}
\bibliography{bibliography.bib}

\end{document}